%% file: 11473.tex
\documentclass[oldversion]{aa}  

\usepackage{natbib}
\bibpunct{(}{)}{;}{a}{}{,}
\usepackage[figuresright]{rotating}
\usepackage{graphicx}
\usepackage{txfonts}

\newcommand{\Msun}{M$_{\sun}$}
\newcommand{\R}{R$_{\textsc{e}}$}

\begin{document}
 
\title{
A New Window of Exploration in the Mass Spectrum:\\
Strong Lensing \emph{by} Galaxy Groups in the SL2S
}
   \titlerunning{Strong Lensing \emph{by} Galaxy Groups}
   \authorrunning{Limousin et~al.}
   \subtitle{}
   \author{M. Limousin\inst{1,2,3},
  R. Cabanac\inst{1},
  R. Gavazzi\inst{4,5},
  J.-P. Kneib\inst{2},
  V. Motta\inst{6},
  J. Richard\inst{7,8},
  K. Thanjavur\inst{9},
  G. Foex\inst{10},\\
  R. Pello\inst{10},
  D. Crampton\inst{11}, 
  C. Faure\inst{12}, 
  B. Fort\inst{4,5},
  E. Jullo\inst{2},
  P. Marshall\inst{13},
  Y. Mellier\inst{4,5},
  A. More\inst{2},
  G. Soucail\inst{10},\\
  S. Suyu\inst{14},
  M. Swinbank\inst{15},
  J.-F. Sygnet\inst{4,5},
  H. Tu\inst{16,4,5},
  D. Valls-Gabaud\inst{17},
  T. Verdugo\inst{6} \&
  J. Willis\inst{9}
         \thanks{Based on observations obtained with MegaPrime/MegaCam, a joint
	 project of CFHT and CEA/DAPNIA, at the Canada-France-Hawaii Telescope
	 (CFHT) which is operated by the National Research Council (NRC) of Canada,
	 the Institut National des Sciences de l'Univers of the Centre National
	 de la Recherche Scientifique (CNRS) of France, and the University of
	 Hawaii. This work is based in part on data products produced at TERAPIX
	 and the Canadian Astronomy Data Centre as part of the Canada-France-Hawaii
	 Telescope Legacy Survey, a collaborative project of NRC and CNRS.
	 Based on observations made with the NASA/ESA Hubble Space Telescope, obtained  
	 at the Space Telescope Science Institute, which is operated by the Association of 
	 Universities for Research in Astronomy, Inc., under NASA contract NAS 5-26555. 
	 These observations are associated with programs 10876 and 11289.}
	 }

   \offprints{marceau.limousin@ast.obs-mip.fr}

   \institute{
	Laboratoire d'Astrophysique de Toulouse-Tarbes, Universit\'e de Toulouse, CNRS,
	57 avenue d'Azereix, 65\,000 Tarbes, France
   	\and
	Laboratoire d'Astrophysique de Marseille, UMR\,6110, CNRS-Universit\'e de Provence, 
	38 rue Fr\'ed\'eric Joliot-Curie, 13\,388 Marseille Cedex 13, France
	\and
        Dark Cosmology Centre, Niels Bohr Institute, University of Copenhagen,
        Juliane Maries Vej 30, 2100 Copenhagen, Denmark
	\and
	CNRS, UMR\,7095, Institut d'Astrophysique de Paris, F-75014, Paris, France
	\and
	UPMC Universit\'e Paris 06, UMR\,7095, Institut d'Astrophysique de Paris, F-75014, Paris, France
	\and
	Universidad de Valpara\'iso, Departamento de F\'isica y Astronomia, Avenida Gran Breta\~na 1111, Valpara\'iso, Chile
	\and
	Durham University, Physics and Astronomy Department, South Road, Durham DH3 1LE, UK
	\and
	Department of Astronomy, California Institute of Technology, 105-24, Pasadena, CA91125, USA
	\and
	Department of Physics and Astronomy, University of Victoria, Victoria, BC, Canada, V8W 3P6
	\and
	Laboratoire d'Astrophysique de Toulouse-Tarbes, Universit\'e de Toulouse, CNRS,
	14 avenue Edouard Belin, 31\,400 Toulouse, France
	\and
	Herzberg Institute of Astrophysics, National Research Council, 5071 West Saanich Road, Victoria, BC V9E 2E7, Canada
	\and
	Laboratoire d'Astrophysique, Ecole Polytechnique F\'ed\'erale de Lausanne (EPFL), Observatoire de Sauverny, CH-1290 Versoix, Switzerland
	\and
	Department of Physics, University of California, Santa Barbara, CA 93106, USA
	\and
	Argelander-Institut f\"ur Astronomie, Universit\"at Bonn, Auf dem H\"ugel 71, 53121 Bonn, Germany
	\and
	Institute for Computational Cosmology, Department of Physics, Durham University, South Road, Durham DH1 3LE, UK
	\and
	Physics Department \&  Shanghai Key Lab for Astrophysics, Shanghai Normal University, 100 Guilin Road, Shanghai 200234,  China
	\and
	Observatoire de Paris, GEPI, CNRS-UMR\,8111, 5 place Jules Janssen, 92195 Meudon Cedex, France
	}

   
  \abstract
   {
   The existence of strong lensing systems
   with Einstein radii covering the full mass spectrum, from $\sim 1-2\arcsec$
   (produced by galaxy scale dark matter haloes)
   to $>10\arcsec$ (produced by galaxy cluster scale haloes) have long been predicted.
   Many lenses with Einstein radii around $1-2\arcsec$ and above $10\arcsec$ have 
   been reported but very few in between. In this article, we present a sample of 13 strong lensing
   systems with Einstein radii in the range $3\arcsec\,-\,8\arcsec$ (or image separations
   in the range $6\arcsec\,-\,16\arcsec$), i.e. systems
   produced by \emph{galaxy group scale dark matter haloes}.
   This group sample spans a redshift range from 0.3 to 0.8.
   This opens a new window of exploration in the mass spectrum, around
   10$^{13}$-\,10$^{14}$\,M$_{\sun}$, a crucial range for understanding the 
   transition between galaxies and galaxy clusters, and a range that have not been extensively probed with lensing techniques.
   These systems constitute a subsample of the Strong Lensing Legacy Survey (SL2S),
   which aims to discover strong lensing systems in the Canada France Hawaii
   Telescope Legacy Survey (CFHTLS). The sample is based on a search over
   100 square degrees, implying a number density of $\sim$\,0.13 groups per
   square degree.
   Our analysis is based on multi-colour CFHTLS images complemented with 
   \emph{Hubble Space Telescope} imaging and ground based spectroscopy.
   Large scale properties are derived from both the light distribution 
   of elliptical galaxies group members and weak lensing of the faint 
   background galaxy population.
   On small scales, the strong lensing analysis yields Einstein radii 
   between 2.5$\arcsec$ and 8$\arcsec$.
   On larger scales, strong lens centres coincide with peaks of 
   light distribution, suggesting that light traces mass. 
   Most of the luminosity maps have complicated shapes,
   implying that these intermediate mass structures may be dynamically young.
   A weak lensing signal is detected for 6 groups and upper limits 
   are provided for 6 others. Fitting the reduced shear with a Singular 
   Isothermal Sphere, we find $\sigma_{\rm SIS}\,\sim$ 500\,km\,s$^{-1}$ with 
   large error bars and an upper limit of $\sim$\,900\,km\,s$^{-1}$ for 
   the whole sample (except for the highest redshift
   structure whose velocity dispersion is consistent with that of a galaxy cluster).
   The mass-to-light ratio for the sample is found to be
   M/L$_i$\,$\sim$ 250 (solar units, corrected for evolution), with an upper 
   limit of 500. This compares with mass-to-light ratios
   of small groups (with $\sigma_{\rm SIS} \sim$ 300 km\,s$^{-1}$) and 
   galaxy clusters (with $\sigma_{\rm SIS} >$ 1\,000 km\,s$^{-1}$), 
   thus bridging the gap between these mass scales.
   The group sample released in this paper will be complemented 
   with other observations, providing a unique sample to study this important 
   intermediate mass range in further detail.
   }

   \keywords{Gravitational lensing: strong lensing --
               Galaxies: groups --
	     }

   \maketitle
 
\section{Introduction}

\subsection{Galaxy Groups \& Cosmology}
Galaxy groups are believed to play a key role in the formation and evolution
of structures in the Universe. They contain the majority of the galaxies
in the Universe, and within a hierarchical framework, they span the regime
between individual galaxies and massive galaxy clusters, making them cosmologically
significant.
They are also more varied in their properties than galaxy clusters, as demonstrated 
by comparisons between various scaling relations in galaxy groups to those
in galaxy clusters.
This indicates that galaxy groups are probably not a homogeneous class of objects,
e.g. they cannot be considered simple scaled-down versions of galaxy clusters. 
Detailed studies of this intermediate regime of the mass spectrum 
($\sim$10$^{13}$-\,10$^{14}$ \Msun) are needed to understand better the physical
processes behind formation of galaxy groups and how galaxy groups
participate to structure formation and evolution.

Properties of low and intermediate redshift galaxy groups ($z<0.5$) have 
been studied using X-ray and optical tracers
\citep[\emph{e.g.}][and references therein]{group1,group2,group3,group4,group5,group6,mamongroup,group7,fabio,group8,group9,sun}.
Galaxy groups have also been studied numerically
\citep[\emph{e.g.}][and references therein]{elena05,jespergroup}.

\subsection{Strong Lens Statistics}
\begin{figure} 
\begin{center}
\includegraphics[scale=0.65]{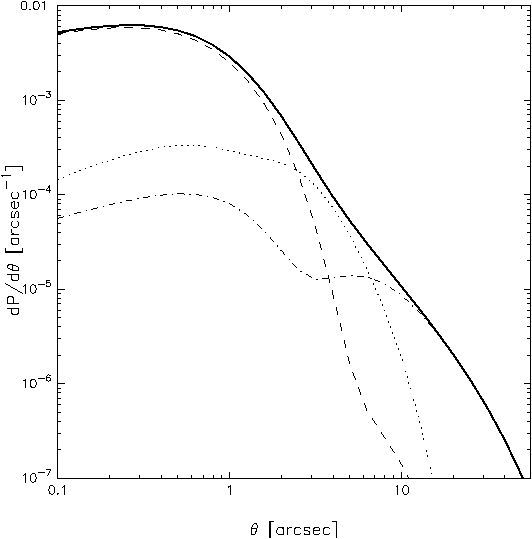}
\caption{The contributions of different types of haloes on the image separation distribution,
from \citet{oguri06}:
galaxy (dashed); groups (dot) and clusters (dot-dashed).
The sum of distributions of three types is shown by thick solid line.}
\label{Fig9oguri06}
\end{center}
\end{figure}

Fig.~\ref{Fig9oguri06} shows the predicted contributions of different types of haloes in the image separation distributions \citep[from][Fig.~9]{oguri06} extracted from N-body simulations, where $\theta$ corresponds to about twice the Einstein radius of the lens\footnote{In this paper, we will use the Einstein radius \R\, to characterise the lenses.}.
This plot shows that there are large overlaps among the halo types, but also that some scales are dominated by different regimes:
on small scales, the distribution is dominated by galaxy-scale dark-matter haloes, with 
\R\,$<$\,2$\arcsec$, and on larger scales, by cluster-scale dark-matter haloes, with
\R\,$>$\,10$\arcsec$.
In between, the distribution is dominated by haloes generating strong lensing deflection of radii between
$\sim$\,3$\arcsec$ and $\sim$\,8$\arcsec$, whose mass is in the range 
10$^{13}$-\,10$^{14}$\,M$_{\sun}$, i.e. by \emph{group-scale dark-matter haloes}.

Hence according to simulations, strong lenses from intermediate mass systems 
should exist in a $\Lambda$CDM Universe. Until now, only a few cases have been detected observationally in the SDSS, which we discuss in Section~\ref{cassowary}.

\paragraph{Strong Lensing Deflector:}
In this paper, we refer to a ``deflector'' as the foreground mass distribution 
giving rise to multiple images, and we focus on its central part. 
Specifically, a strong lensing deflector refers to the area of the
foreground mass distribution enclosed by the multiple images. 
This region can be populated by more than a single galaxy on scales smaller 
than \R, as it is the case for some of the galaxy groups reported in this work.
In order to avoid confusion, we use the term deflector instead of lensing galaxy.
This is particularly important for group scale deflectors since their mass may
no longer be associated to a single galaxy.
This definition is well adapted to most of the groups studied in this paper,
where the multiple images identification is clear and therefore the region enclosed
by these images is well defined.
However, this is not the case for three of the groups presented in this paper.

\subsection{Strong Lensing \emph{in} and \emph{by} a Galaxy Group}
We shall emphasise the distinction between strong lensing \emph{in} a galaxy group
and \emph{by} a galaxy group.
Strong lensing in a galaxy group corresponds to the observation of a
strong lensing feature, with a typical Einstein radius of $\sim2\arcsec$,
generated by a galaxy group member. Since the high-density environment is likely to
enhance the strong galaxy-galaxy lensing cross section \citep[see][but see \citet{cecile} in the \textsc{cosmos} field]{kovner,ole,fassnacht06,oguri05,king1689,slacsclustering},
many strong lensing systems in groups (or clusters) have been reported
\citep[\emph{e.g.}][]{kundic97a,fassnacht02,morgan05,williams06,momcheva,auger07,ring1689,auger08,grillo12,treuaroundslacs,inada}.
If the creation of multiple images is boosted by external shear and convergence from a 
smooth group component, these strong lensing systems are, to first order, generated by a
galaxy-scale dark-matter halo populated by a single galaxy \citep[most of the time an 
elliptical galaxy, but note that efforts are currently underway to find lenses generated 
by haloes populated by spiral galaxies, see][]{chloe,trott}.

On the other hand, strong lensing by a galaxy group corresponds to the observation of
strongly lensed features with a typical Einstein radius between $\sim 3\arcsec$ and
$\sim 8\arcsec$. This translates into a projected mass enclosed within this radius
around $\sim$ 10$^{12.5}$\,-\,10$^{13}$\,\Msun, i.e. characteristic of a 
\emph{group-scale dark-matter halo}. Note that the highest limit of 8$\arcsec$ 
adopted here is somewhat arbitrary and may be considered by others
as a characteristic size of a poor cluster. 

\subsection{Galaxy Groups in the Strong Lensing Legacy Survey}

A straightforward method for finding lens systems is to identify them 
directly within existing multi-band wide-field imaging surveys.
For this purpose, we have started to explore systematically the Canada France Hawaii 
Telescope Legacy Survey (CFHTLS),
using a dedicated automatic search procedure optimised for the detection of arcs.
This subsample of the CFHTLS forms the Strong Lensing Legacy Survey \citep[SL2S,][]{sl2s}.
This sample is not biased towards high masses by any prior X-ray selection and because
of the wide area covered, cosmic variance is neglected and the resulting sample of 
strong lenses is assumed to be representative of what can be found in the Universe.
However, selection biases inherent to our detection technique do exist and will be 
addressed in a forthcoming publication using numerical simulations.
The SL2S sample probes a wide range of masses and a wide range of redshifts for the lenses.
Thanks to unsurpassed combined depth, area and image quality of the CFHTLS,
we have uncovered a new population of lenses, with Einstein radii between $\sim 3\arcsec$ 
and $\sim 8\arcsec$, i.e. generated by \emph{galaxy-group scale-dark matter haloes}.
This new population effectively bridges the gap between single galaxies 
and massive clusters, opening a \emph{new window of exploration in the mass spectrum}.
The purpose of this paper is to present the first representative sample of such lenses.
Note that this sample is not statistically complete since the survey is not complete yet
and that we do not understand properly our selection function.

This paper is organised as follows:
data used in this work are presented in Section 2.
Sections 3, 4 and 5 present the methodology of the systematic analysis we
apply on each group.
Results are presented in Section 6 and discussed in Section 7.
In the Appendix, each group is described individually, with some images.
All our results are scaled to a flat, $\Lambda$CDM cosmology with $\Omega_{\rm{M}} = 0.3, \
\Omega_\Lambda = 0.7$ and a Hubble constant \textsc{H}$_0 = 70$
km\,s$^{-1}$ Mpc$^{-1}$.
All images are aligned with WCS coordinates, i.e. north is up, east is left.
Magnitudes are given in the AB system.

\section{Data}
In this Section, we present the various data sets used in this work.
Multicolour imaging from the CFHTLS constitutes the basis of our analysis and is 
complemented with ongoing HST imaging and ongoing ground-based spectroscopy.

\subsection{The Canada-France-Hawaii Telescope Legacy Survey}
\subsubsection{Description}
The CFTHLS is a major photometric survey of more than 450 nights over 5 years
(started on June 1st, 2003, ending 2008) using the wide field imager 
\textsc{MegaPrime} which covers $\sim$\,1 square degree on the sky, with a pixel size of
0.186$\arcsec$. The project is comprehensively described in 
{\tt http://www.cfht.hawaii.edu/Science/CFHLS/} and links therein. The CFHTLS
has two components aimed at extragalactic studies: a very Deep component
made of 4 pencil-beam fields of 1\,deg$^2$ and a Wide component made of
3 mosaics covering 170\,deg$^2$ in total, both in 5 broadband filters. The
data are pre-reduced at CFHT with the Elixir pipeline\footnote{http://www.cfht.hawaii.edu/Instruments/Elixir/} 
which removes the
instrumental artifacts in individual exposures. The CFHTLS images are
then evaluated, astrometrically calibrated, photometrically 
inter-calibrated, resampled and stacked by the Terapix group at the 
Institut d'Astrophysique de Paris (IAP) and finally archived at the Canadian 
Astronomy Data Centre (CADC). Terapix also provides weightmap images,
quality assessments meta-data for each stack as well as mask files 
that mask straylight, saturated stars and defects on each image.

\subsubsection{Terapix release T0004}
The SL2S group sample presented here is based on the T0004 release
(July 2007), corresponding to data obtained between the Spring 2003
and the spring 2007.  A detailed description of this release
is given at the Terapix web site\footnote{http://terapix.iap.fr} and
CADC site\footnote{http://www2.cadc-ccda.hia-iha.nrc-cnrc.gc.ca/cfht/T0004.html}. 
The T0004 release includes 120 deg$^2$ of stacked fields in the Wide survey 
observed in broadband $g'$, $r'$ and $i'$ filters, and a stack of the 
35 Wide and 4 Deep fields in the 5 bands, for a total area of 124 deg$^2$, or 
ca. 110 deg$^2$ of unmasked area, i.e. area not contaminated by instrumental 
artefacts from bright stars (internal reflections, bleeding).
Table~\ref{cfhtls_t4} summarises the main characteristics of the T0004 release.
The 4 Deep fields are much deeper, with a seeing 
ranging from 0.9\arcsec\ in $u^*$ to 0.7\arcsec\ in $z'$. 
Because of the observing strategy, the Wide survey image quality is prone 
to large variations from 1\arcsec to 0.6\arcsec seeing in all bands.
Nevertheless, the Wide survey is more suited to our 
strong lensing selection processes, because of its wide angular
coverage.

\begin{table}
\caption{CFHTLS: Terapix T0004 release (July 2007).
{\bf{
For a point source $S/N = 5$ ($AB$ to Vega $u^*-0.35$, $g'+0.09$, $r'-0.17$, $i'-0.40$, $z'-0.55$)}}}
\label{cfhtls_t4}
\begin{tabular}{lccccc}
\hline
{\bf Deep fields} & \multicolumn{5}{c}{\bf AB Magnitudes Limits}\\
& $u^*$&$g'$&$r'$&$i'$&$z'$\\
\noalign{\smallskip}
D1-25 & 26.7 & 27.5 & 27.5 & 27.3 & 26.0\\
D2-25 & 25.6 & 27.3 & 27.3 & 27.0 & 25.6\\
D3-25 & 26.6 & 27.5 & 27.4 & 27.2 & 26.1\\
D4-25 & 26.9 & 27.5 & 27.5 & 27.1 & 26.0\\
D1-25 seeing & 0.93 & 0.93 & 0.75 & 0.74 & 0.71\\
D2-25 seeing & 0.83 & 0.83 & 0.77 & 0.71 & 0.72\\
D3-25 seeing & 0.89 & 0.89 & 0.77 & 0.73 & 0.64\\
D4-25 seeing & 0.86 & 0.86 & 0.77 & 0.72 & 0.71\\

\noalign{\medskip}
{\bf Wide fields} & \multicolumn{5}{c}{\bf AB Average Magnitudes Limits}\\
\noalign{\medskip}
W1  & 25.8 & 26.5 & 26.0 & 25.8 & 25\\
W2  & 25.8 & 26.5 & 26.0 & 25.8 & 25\\
W3  & 25.8 & 26.5 & 26.0 & 25.8 & 25\\
W4  & 25.8 & 26.5 & 26.0 & 25.8 & 25\\
\noalign{\smallskip}
W1 area (unmasked) & \multicolumn{5}{c}{44 (39) deg$^2$}\\
W2 area (unmasked) & \multicolumn{5}{c}{~20 (~16) deg$^2$}\\
W3 area (unmasked) & \multicolumn{5}{c}{40 (~34) deg$^2$}\\
W4 area (unmasked) & \multicolumn{5}{c}{16 (~13) deg$^2$}\\
\noalign{\smallskip}
\hline
\end{tabular}
\end{table}

\subsection{Space Based Imaging}
The strong lensing features detected from ground-based images have been observed with the
\emph{Hubble Space Telescope} (HST) for 11 of the 13 groups presented in this paper.
Observations were done in snapshot mode (C\,15 and C\,16, P.I. Kneib, ID 10876 and 11289).
3 groups have been imaged in three bands with the ACS camera (F814, F606 and F475), 
and the remaining in the F606 band only with the WFPC2 camera.
Space-based imaging generally allows one to better resolve the strong lensing systems
we are interested in.

\subsection{Spectroscopy}
Various ongoing spectroscopic campaigns are targeting both the galaxy group members 
as well as the multiply imaged systems.
We use the following facilities:
\begin{enumerate}
\item FORS\,2 on VLT: resolution of 600\,RI, exposure
2\,800 s, with $\Delta \Lambda$ from 5\,000 to 8\,000 \AA.
\item LRIS on Keck: Dichroic 680 and 560 nm, exposure 600s and 1200s with a slit of 1.5$\arcsec$.
\item Gemini (Program GN-2007A-Q-114):
we observed SL2S\,J14300+5546 on June 2007 using GMOS LS+R400 grating at 7500 \AA\ during 3\,750 s.
\end{enumerate}
When available and relevant to this work, we will report some of the results 
of our spectroscopic campaigns.
Details of the observing runs and spectra will be presented in forthcoming publications.
For SL2S\,J14081+5429 and SL2S\,J22214-0053, we report the spectroscopic redshift 
measured by the Sloan Digital Sky Survey (SDSS).

\subsection{Photometric Redshifts}
Photometric redshifts were estimated for all the groups, based on the magnitudes of the
brightest galaxy populating the strong lensing deflector whose coordinates are given in
Table~\ref{res}.
Aperture magnitudes were extracted with \textsc{sextractor} 2.5 \citep{sextractor} and
photometric redshifts were estimated using the HyperZ software \citep{hyperz}.
For all groups but two (SL2S\,J14300+5546 and SL2S\,J14314+5533), 5 photometric bands 
are available.
Photometric redshifts, with 3$\sigma$ error bars are given in Table~\ref{res}. They can
be compared to the spectroscopic measurements available for 9 of the groups.
We note that in the case of 4 groups over 9, the spectroscopic redshifts are not consistent
with the 3$\sigma$ confidence intervals of the photometric redshifts estimates.
However, the difference between each estimate is smaller than 0.03. This has no influence
on the lensing properties of the galaxy groups and therefore does not bias our analysis.

In addition, we have been looking systematically in the NED database at
all known sources within 5$'$ of each strong lensing event to gather 
any observational data relevant to this study.

\section{Building Luminosity Maps}
Luminosity maps were created from the $i$-band isophotal magnitudes of all the 
elliptical galaxies tagged as members of the group.
This is the deepest band that best represents the old stellar population of
elliptical galaxies.
Catalogues of objects positions and magnitudes have been generated using 
\textsc{sextractor} 2.5.

\subsection{Selecting Group Members}
We build matched $g,r,i$ bands catalogues of all objects with a radius of 5$'$ of the
lens centre. We compute the $r-i$ colour of the brightest galaxy populating 
the strong lensing deflector. We refer to this quantity in the following as $(r-i)_{\,0}$. 
We then consider all galaxies whose $r-i$ colour satisfy:
\begin{equation}
(r-i)_{\,0} -\,0.15 < r-i <\, (r-i)_{\,0} +\,0.15\,\,\&\,\, i < {\rm mag}_{\rm lim} (z)
\end{equation}
as possible galaxy group red-sequence members \citep[\emph{e.g.}][]{gladders}.
We miss the faintest group members but the most luminous ellipticals 
are correctly selected.
When spectroscopy of group members is available, we verified that our procedure 
does select the spectroscopically confirmed group members.
A more robust selection will require us to analyse multi-object spectroscopy 
of the fields.

In order to compare the group luminosity maps across the redshift range, 
we restrict our group member catalogue to galaxies brighter than a
magnitude threshold ${\rm mag}_{\rm lim}$($z$).
This magnitude threshold is set by the apparent magnitude of the faintest 
elliptical member of the highest redshift group (SL2S\,J22133+0048 at 
$z_{\rm phot}=0.83$).
That apparent magnitude is converted into a rest-frame absolute 
luminosity, converted back into a redshift-dependent apparent magnitude 
mag$_{\rm lim}$($z$) for each group. 

\subsection{Contamination}
The red sequence technique is known to suffer from contamination.
In our case, because the number count of objects per group that satisfy Equ.~1 
is small (less than 100), we visually inspected each group member 
candidate and manually rejected galaxies that were unlikely
to be physically connected to the group (e.g. higher redshift spiral galaxies).
We also estimated contamination from background/foreground objects serendipitously
falling in the same colour bin from random fields, i.e. fields large 
enough to ensure that they are, on average, empty.
Using CFHTLS deep fields galaxy catalogues and applying Equ.~1 
selection, we estimated the total background/foreground luminosity that 
adds up to the group catalogues. 
This estimation is consistent with what we have manually removed from
our visual inspection. Therefore, we prefer removing the background 
contamination by eye since the method allows us to identify the 
contaminating galaxies, whereas a statistical approach would lead to
subtraction of constant luminosity sheets across groups 
and a loss of positional information.

\subsection{Luminosity Contours}
We build luminosity maps to help us characterise the large-scale luminous
properties of the group sample.
The CFHTLS $i$-band image is divided into cells 20-pixels wide (corresponding to 3.7$\arcsec$).
Considering the centre of each cell, 
we compute the total rest-frame $i$-band luminosity $L$ by summing 
individual luminosities of the five closest group members, deduced from their 
isophotal magnitude $M$:
\begin{equation}
L = 10^{ ( M_{\sun} - M + \rm{DM} + k + ev) / 2.5}
\end{equation}
where $M_{\sun}$ is the Sun $i$-band absolute magnitude; DM is the distance modulus,
k the k-correction factor and ev the passive evolution correction, 
estimated from an elliptical template \citep{BC03}.
This total luminosity is divided by the circular area enclosing the fifth neighbour:
this gives a density within the considered cell that we express in
units L$_{\sun}$\,kpc$^{-2}$.
The computed luminosity density map is then convolved with a Gaussian kernel of 
width 3 cells.

The choice for 5 neighbours is small compared to the number of catalogued group members 
(on average 46 members).

We found that the size of the cells does not influence the shape of
the resulting luminosity contours. We checked that the distance to 
the fifth neighbour is always larger than cell/$\sqrt 2$, which means 
we do not undersample any region of the field.

\section{Weak Gravitational Lensing Analysis}

We perform a one dimensional weak lensing (WL) analysis on each group
with the goal of probing the projected mass distribution on large scales. 
This technique is well established and is widely used on galaxy cluster scales
\citep[\emph{e.g.}][]{bardeau07,hoekstrasample}.
The basic idea is the following:
from the shape of the background lensed galaxy population, we
can estimate the shear vector \citep[or more precisely, the reduced shear, see 
\emph{e.g.}][]{mellier99}.
Its tangential component (with respect to the centre of the deflector) is 
directly proportional to the mass distribution of the deflector,
whereas its radial component is expected to be equal to zero in the case of 
perfect data.
Given the relatively smaller mass of galaxy groups
(compared to galaxy clusters), detecting weak lensing by groups of galaxies is 
challenging.
Until now, weak lensing signals of such mass scales has been recovered only 
by stacking groups \citep{hoekstra01,parker05,rachelgroup}.
Moreover, the strength of the weak lensing signal is expected to decrease 
with the redshift of the deflector, thus we expect to detect a weak lensing 
signal only for the lower redshift groups.
In any case, an upper limit on the velocity dispersion of the structures 
studied in this work is of interest.
Because strongly lensed source redshifts are unknown, weak-lensing derived 
velocity dispersion upper limits constitute the only way to show that 
our sample contains only groups or poor clusters and no massive clusters.
Indeed, an Einstein radius of $\sim 6\arcsec$ could be generated by a massive 
cluster lensing a low redshift background source.

\subsection{Weak Lensing Methodology}
We have been following the methodology developed by \citet{bardeau05,bardeau07},
using the \textsc{sheartool} package \citep{sebphd}.
We refer the reader to these papers for a detailed description of the 
cataloguing.
Here we give a brief outline of the different
steps involved in the analysis of the reduced and calibrated images.

The first step is to construct a photometric catalogue for each individual 
image. Object positions and magnitudes were extracted with 
\textsc{sextractor} 2.5. The second step of the analysis is to
extract a star catalogue which will be used to estimate the local Point 
Spread Function (\textsc{psf}). We select stars and clean the resulting 
catalogue as described in \citet{bardeau05}. In order to
measure the shapes of the stars, we used the \textsc{im2shape} software
developed by \citet{im2shape}. At this stage, we have a map of the
\textsc{psf} distribution over the entire field. The third step is to compute 
the galaxy catalogues selected for the weak lensing analysis. 
Galaxies are selected from the photometric catalogues
according to the criterion described in \citet{bardeau05}. To
measure the shapes of galaxies, we first linearly interpolate the
local \textsc{psf} at each galaxy position by averaging the shapes of the five
closest stars. The number of 5 stars is found to be large enough to
locally interpolate the \textsc{psf}, whereas choosing a much larger number 
would over-smooth the \textsc{psf} characteristics. \textsc{im2shape} then 
computes intrinsic shapes of galaxies by convolving a galaxy model with the
interpolated local \textsc{psf}, and determines which one is the most likely by
minimising residuals. In the end, \textsc{im2shape}'s output gives a
most likely model for the fitted galaxy characterised by its position,
size, ellipticity and orientation, and errors on all of these quantities.
For the purpose of the weak lensing analysis, we use the shape parameters 
derived from the CFHTLS $i$ band: observations in this photometric band have 
been optimised for weak lensing purposes, thus data quality in terms of 
seeing and source density is superior to the other bands.

Note that the reliability of the \textsc{im2shape} software
has been validated through the simulated data of the STEP project 
\citep{step1} and the results show that no significant bias is introduced 
by the deconvolution.
It has been successfully applied in a number of weak lensing works
\citep{kneib03,edu,edu05,bardeau05,bardeau07,mypaperII,mypaperIII,elinor,edu07}.

\subsection{Background Galaxies: Selection and Redshift Distribution}
We select as background lensed sources the galaxies whose $i$-band magnitudes 
fall between 21.5 and 24, hence close to the completeness magnitude for all 
groups (Table~\ref{res}).
The mean density of background galaxies equals to $\sim$\,12 per square arcminute.
Applying the same magnitude cuts to the four CFHTLS deep fields catalogues, we find
a density of $\sim$\,15 per square arcminute, slightly larger than the mean density of 
background galaxies used in our weak lensing analysis.
This is expected since we have kept only objects whose shape parameters
are reliably measured, as detailed in \citet{bardeau05,bardeau07}.

In order to relate the strength of the weak lensing signal to a physical
velocity dispersion characterising the group potential, one needs to estimate 
the mean geometrical factor D$_{\rm ls}$/D$_{\rm s}$, 
where D$_{\rm ls}$ is the angular diameter distance between the lens (here the 
group) and the source and D$_{\rm s}$ is the angular diameter distance between 
the observer and the source. For this estimation, we consider the
photometric redshift catalogue from the T0004 release of the
CFHTLS-Deep survey, which corresponds to the deep stacks of the repeated 
observations of four independent \textsc{MegaCam} fields.
These data are therefore collected with the same instrument in the same 
photometric bands as the data used in this work. The CFHTLS-Deep is much deeper 
(at least one magnitude deeper in $r$) and the multicolour observations in 5 
bands allow the determination of photometric redshifts.
We have used the publicly available photometric redshift catalogue provided by 
Roser Pello\footnote{http://www.ast.obs-mip.fr/users/roser/CFHTLS\_T0004/} which 
has been calibrated and validated with spectroscopic samples \citep{iena}.
We applied to this catalogue the same magnitude selection criteria that we applied to our 
background galaxy catalogue (i.e. $i$ between 21.5 and 24).
For each group, we compute the average geometrical factor 
D$_{\rm ls}$/D$_{\rm s}$ by integrating this redshift probability distribution 
between 0 and 5.
We also compute the effective source redshift $z_{\rm eff}$ that is defined by
the redshift at which the geometrical factor becomes equals to the mean geometrical
factor. Both quantities are given in Table~\ref{res}.

\citet{coupon} recently estimated photometric redshift for 35 deg$^2$ of the 
Wide survey. Since our group catalogue is based on a search of 100 deg$^2$, all 
groups do not fall in the study by \citet{coupon}. Moreover, they provide 
photometric redshifts for galaxies brighter than $i=22.5$ whereas our background 
galaxy catalogues contains objects as faint as $i=24$.
Therefore, we cannot benefit from this work in assigning photometric redshifts 
individually to each background galaxy.

\citet{coupon} also estimate photometric redshifts for the Deep fields down 
to $i=24$.
A thorough comparison of both catalogues is beyond the scope of this paper, but we can
check if the presented results change significantly if we use this catalogue
instead of ours. We apply the same magnitude selection to the \citet{coupon} catalogue
and recompute the quantities of interest.
The main quantity we are interested in is the weak lensing inferred mass.
Therefore, we present in Table~\ref{res} in brackets the masses derived using the
\citet{coupon} catalogue that can be compared to the masses derived using our catalogue. 
We find the difference between each estimate to be much
smaller than the associated error bars.
Therefore, we conclude that adopting one photometric catalogue or
the other does not change significantly the presented results.

\subsection{Fitting the Shear Profile}
For each group, we look for a shear signal centred on the lens
between 150\,kpc and 1.2\,Mpc from the lens.
In the case of SL2S\,J08544-0121, a bimodal group, we use the barycentre of the
light distribution instead since the lens does not dominate the light 
distribution.
We fit an Singular Isothermal Sphere model (SIS) to the signal within this range.
The choice of this model is motivated by the fact that we do not expect to get a
weak lensing signal strong enough that could allow us to probe more complicated 
mass profiles. We use the average geometrical factor to convert the fitted 
Einstein radius into an SIS velocity dispersion $\sigma_{\rm SIS}$.

\section{Strong Gravitational Lensing Analysis}
The strong lensing features can be divided in two categories.
Of the 13 groups, 8 exhibit a typical 'cusp' configuration where a multiply 
imaged system forms a characteristic arc on one side of the deflector, sometimes 
with a counter image on the other side of the deflector.
The other 5 groups have more complex strong lensing configurations for which 
identification of multiple images is challenging and for which no strong lensing 
modelling is possible at this stage. 
More details on each strong lensing configuration are discussed in the Appendix.

For the 8 cusp groups, we apply a simple mass modelling in order to estimate 
the Einstein radius, using the \textsc{lenstool} code \citep[see][for a 
description of the method and how error bars are derived]{jullo07}.
The refinement of this modelling will depend on the number of available 
observational constraints.
A realistic parametric mass modelling requires at least 5 parameters. 
For instance a Singular Isothermal Ellipsoid (SIE) has position (x,y), 
ellipticity ($e$), position angle (PA) and Einstein radius.
For a multiply imaged system where a background source is imaged $n$ times, 
and assuming its redshift is known, we have 2\,$\times$\,($n$-1) constraints. 
Only if $n \geq$\,4 can we constrain an SIE profile.
In the cusp sample, we have typically only one multiply imaged
system, composed of 3 to 4 images which translates into 4 to 6 observational 
constraints.
When the number of observational constraints does not allow to probe 5 
free parameters, we set the ellipticity and position angle to 0, leading to a 
non-elliptical mass distribution.
The main robust quantity we aim to derive is the Einstein radius. 
For the non-cusp group, we evaluate the Einstein radius simply by measuring 
the distance between the strong lensing feature and the brightest group galaxy.

Optimisations are performed in the image plane, assuming a positional uncertainty
equals to 0.1$\arcsec$.
The ellipticity of the mass distribution was forced to be smaller 
than 0.6 (in units $(a^2-b^2)/(a^2+b^2)$) as motivated by numerical simulations 
\citep{jingsuto}.
Table~\ref{slres} summarises information relevant to strong lensing 
modelling: number of constraints available; optimised parameters (position, 
ellipticity, position angle, Einstein radius); RMS in the image plane and 
reduced $\chi^2$, which quantifies the goodness of the fit.
We note that for two systems, we are unable to correctly retrieve the
observational constraints, with RMS larger than the positional uncertainty:
SL2S\,J08544-0121, RMS\,=\,0.28$\arcsec$ and SL2S\,J02215-0647, RMS\,=\,0.2$\arcsec$.

\begin{table*}
\begin{center}
\begin{tabular}{lcccccccc}
\hline
SL2S Group & Nb & X [$\arcsec$]& Y [$\arcsec$] & $e$ & $\theta$ (degree) & R$_{\textsc{e}}$ & RMS & $\chi^2$ \\
\hline
J22214-0053 & 4 & -0.2$\pm$0.06 & -2.0$\pm$0.15 & [0] & [0] & 6.83$\pm$0.15 & 0.08 & 0.0 \\
\hline
J08544-0121 & 8 & 0.0$\pm$0.1 & 0.0$\pm$0.06 &  0.59$\pm$0.0  &  20.8$\pm$1.2 &  5.48$\pm$0.12 & 0.28  &  17\\
\hline
J09413-1100 & 4 & -1.3$\pm$0.2 & -1.9$\pm$0.9  &  [0] & [0]  &  7.50$\pm$0.81  & 0.07 & 1.5 \\
\hline
J14300+5546 & 4 & -0.2$\pm$1.4  & -1.1$\pm$1.5 & [0] & [0] &  4.69$\pm$1.43 & 0.02 & 0.1 \\
\hline
J02140-0532 & 8 & 0.85$\pm$0.3 & -2.6$^{+1.2}_{-0.03}$ & 0.56$\pm$0.11 & 111.2$\pm$0.1 & 7.31$\pm$0.49 & 0.1 & 1.1\\
\hline
J02254-0737 & 4 & -1.9$\pm$0.15 & -2.6$\pm$0.6 & [0]  & [0] & 6.00$\pm$0.31  & 0.04 & 0.4 \\
\hline
J02215-0647 & 6 & 0.14$\pm$0.15 & -1.3$\pm$0.2 & 0.57$\pm$0.01 & 104$\pm$3 & 2.56$\pm$0.14 & 0.2 & 20 \\
\hline
J\,22133+0048 & 4 & -1.7$\pm$0.15 & -0.8$\pm$0.3 & [0] & [0] &  3.40$\pm$0.2 & 0.01 & 0 \\
\hline
\end{tabular}
\end{center}
\caption{Summary of the strong lensing modelling. 
{\bf{Number of constraints Nb and optimised
SIE parameters. Error bars represents 1$\sigma$ confidence level on
the parameters inferred from the \textsc{mcmc} optimisation. Values into brackets are not
optimised. This corresponds to models where the number of observational constraints is smaller than the
number of free parameters characterising the SIE profile. The goodness of the fit is quantified by the
RMS in the image plane and the reduced $\chi^2$.
$e$ is the ellipticity of the mass distribution (in units $(a^2-b^2)/(a^2+b^2)$).
}}}
\label{slres}
\end{table*}

\section{Results}
\label{overview}
\begin{figure*}[h!]
\begin{center}
\includegraphics[scale=1.15]{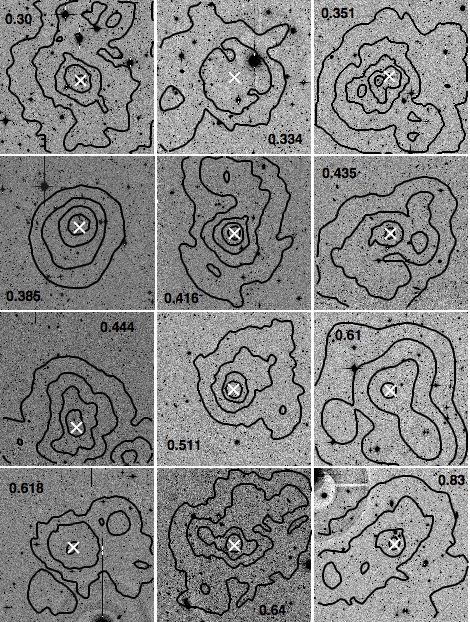}
\caption{Large scale $10'\times10'$ CFHTLS $i$ band image for all groups but 
one (SL2S\,J08591-0345 for which data are missing). 
Picture are ranked by increasing redshifts as in Table~\ref{res}.
Redshift is reported on each picture, with three digits precision when
spectroscopically measured. The white cross shows where the lens is located.
Black contours correspond to luminosity contours (corrected for passive 
evolution) equal to 10$^{5}$, 3$\times$10$^{5}$, 10$^{6}$, 3$\times$10$^{6}$ and 
10$^{7}$ L$_{\sun}$\,kpc$^{-2}$.
}
\label{all}
\end{center}
\end{figure*}

The main properties of the groups are summarised in
Table~\ref{res} and illustrated in the following figures.
The name of each group corresponds to the equatorial coordinates (J\,2000) 
of the brightest member. In the Appendix, we dedicate a paragraph to detail the properties of each group together with the following images:
\begin{enumerate}
\item A large scale $10'\times10'$ CFHTLS $i$ band image.
The white cross shows where the lens is located.
We draw luminosity contours (corrected for passive evolution) equal 
to 10$^{5}$, 3$\times$10$^{5}$, 10$^{6}$, 3$\times$10$^{6}$ and 10$^{7}$ 
L$_{\sun}$\,kpc$^{-2}$.
\item A 1-square-arcminute colour image from $g,r,i$ CFHTLS imaging centred on 
the deflector.
\item The HST image (when available) centred on the deflector, with the image 
size given in the figure caption. The image is in colour when  observations are 
available in three bands (F475, F606 and F814), and is in F606 otherwise.
\end{enumerate}
Some of these visual information are summarised in Fig.~\ref{all}, where we show 
the large scale CFHTLS images for the 12 groups (all except SL2S\,J08591-0345 
for which data is incomplete, see Section~\ref{0859}).

\subsection{Einstein Radii \& Redshift Distributions}
The group sample presented here spans $z = 0.30$ to $z = 0.83$,
with a mean redshift $<z>=0.50$.
We report in Table~\ref{res} the fitted \R\, for the 8 cusp groups, as well as 
the measured \R\, for the remaining 5 non cusp groups. They span 2.5$\arcsec$
to 8$\arcsec$.
We expect the distribution of \R\, to decrease with increasing \R\ (Fig.~1).
This is not seen here, but we need to correct this distribution for selection 
biases. Clearly we will miss more \R$\,\sim2\arcsec$ than \R$\,\sim5\arcsec$ 
systems with our detection robot (simply because a small \R\,is more likely to 
be embedded within the light distribution of the lensing galaxy and missed for that reason).

\subsection{Luminosities}
\begin{figure}
\begin{center}
\includegraphics[scale=0.6]{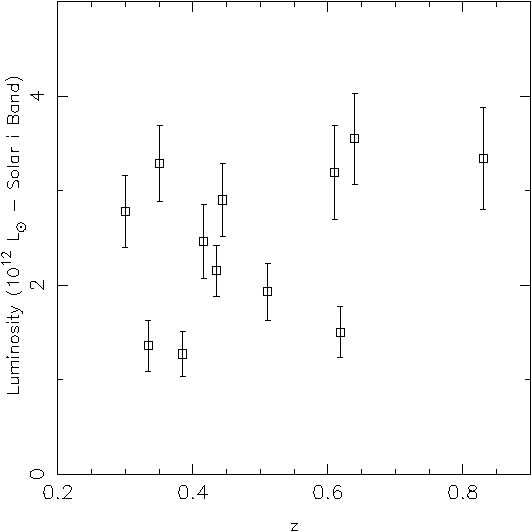}
\caption{
Luminosities (corrected for passive evolution) versus group redshifts.
}
\label{reslum2}
\end{center}
\end{figure}

In all cases, we find that the brightest galaxy populating the strong lensing 
deflector is the brightest group member. Total luminosities, corrected for 
passive evolution, are reported in Table~\ref{res}. 
Fig.~\ref{reslum2} shows the luminosity as a function of redshift.
We find that the group sample presented in this work is homogeneous in terms of
total luminosities.

\subsection{Weak Lensing}

\begin{figure}\begin{center}
\includegraphics[scale=0.6]{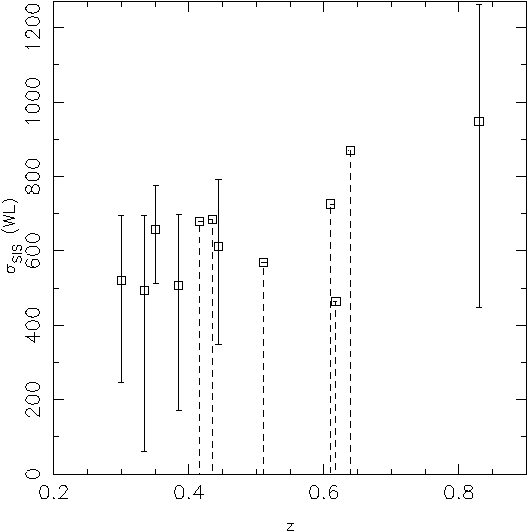}
\caption{
Velocity dispersions inferred from fitting a Singular Isothermal Sphere to the
reduced shear profile, as a function of group redshifts. Dashed
lines corresponds to upper limits only.
}
\label{reswl}
\end{center}
\end{figure}

We report the inferred SIS velocity dispersion for each group in 
Table~\ref{res}.
Fig.~\ref{reswl} shows the fitted $\sigma_{\rm SIS}$ as a function of redshift.
As expected, we see that the weak lensing constraints are tighter for the lowest 
redshift groups.
Considering the whole sample, we see that $\sigma_{\rm SIS} 
\sim$ 500\,km\,s$^{-1}$ with large error bars and an upper limit of 
$\sim$\,900\,km\,s$^{-1}$. Therefore, we verify that
no massive cluster with $\sigma_{\rm SIS} > 1000$ km\,s$^{-1}$, which would 
lens a close background source, is included in our sample and that it 
consists of groups or poor clusters only.
Note, however, that the highest redshift structure included in the present
work has a velocity dispersion between 432 and 1218 km\,s$^{-1}$.
The upper limit is therefore consistent with a galaxy cluster, see Section~\ref{highz}.

We are not able, given the data at hand, to pursue a more refined 2-D 
weak lensing study from which we could infer the shape of the mass 
distribution over the field and compare more precisely with the luminosity 
contours. Indeed, it is already challenging to infer an SIS velocity dispersion 
from our 1-D analysis and error bars are large. A 2 dimensional analysis would 
involve more free parameters and is not feasible here.

\subsection{Mass-to-Light Ratios}
\begin{figure}\begin{center}
\includegraphics[scale=0.6]{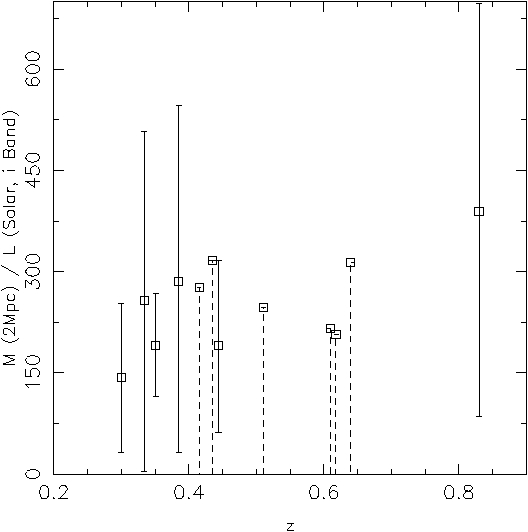}
\caption{
Mass to Light ratio for the group sample, taking into account passive evolution
correction.
Dashed lines corresponds to upper limits only.
}
\label{msurl}
\end{center}
\end{figure}

From the above, we are able to compute mass-to-light ratios.
From the weak lensing analysis, we compute the projected mass of the group within a
circular aperture of 2\,Mpc centred on the lens (except SL2S\,J08544-0121, 
a bimodal group that we centre on the barycentre of the light distribution).
Then we consider the total luminosity, corrected for passive evolution.
Values are given in Table~\ref{res} and illustrated in Fig.~\ref{msurl}.
The mean value is equal to $\sim$\,250 ($i$ band, solar units, corrected for 
passive evolution), with an upper limit of 500.

\subsection{Trends with Luminosity}
\begin{figure}\begin{center}
\includegraphics[scale=0.6]{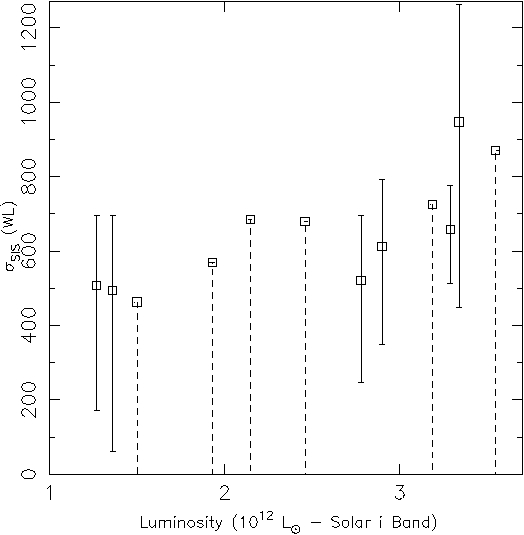}
\caption{
SIS velocity dispersion inferred from the weak lensing
analysis as a function of luminosity (corrected for passive evolution).
}
\label{lumsigma}
\end{center}
\end{figure}

If the mass is proportional to the luminosity of the systems, we expect the SIS 
velocity dispersion to correlate with the luminosity of the galaxy groups.
We show in Fig.~\ref{lumsigma} velocity dispersions inferred from the weak 
lensing analysis as a function of luminosities (corrected for passive 
evolution). The expected trend is suggested but not statistically 
significant due to the large error bars.

\subsection{Number Density}
We have found 13 group scale lenses within a field of $\sim$ 100 square degrees 
(Table~1).
This gives a rough estimate of the number density of $\sim$ 0.13 groups per 
square arcminute. We note that some group candidates found on the CFHTLS images 
were not confirmed on the HST imaging and not included here.
Moreover, at this stage of the project, we do not understand properly our
selection function. Selection biases inherent to our detection technique exist 
and will be addressed in a forthcoming publication using numerical simulations.
Once the SL2S will be completed and the selection function understood, we will 
be able to compare the distribution of observationally determined \R\, to the 
predictions from \citet{oguri06}, which may provide an interesting test of the 
$\Lambda$CDM scenario.

\subsection{The Case of SL2S\,J08591-0345}
\label{0859}
This system at $z_{\rm spec}=0.647$ is presented in the Appendix in more detail.
The observational setup makes this system special: the lens is located close 
(75$\arcsec$\,=\,508\,kpc) to the edge of the field of view. Therefore, it is 
very likely that we are missing a significant part of this
interesting structure (note the exotic strong lensing configuration).
Therefore, we do not report a value of the luminosity for this group, neither do 
we pursue any weak lensing analysis.

\subsection{The Case of SL2S\,J22133+0048}
\label{highz}
The weak lensing analysis of this system located at $z_{\rm phot}=0.83$
yields a velocity dispersion between 432 and 1218 km\,s$^{-1}$ which
makes it consistent with a galaxy cluster.
We note that its Einstein radius is rather small for a galaxy cluster, about 3.4$\arcsec$.
It is possible that SL2S\,J22133+0048 is indeed a galaxy cluster with a velocity dispersion
larger than 1000 km\,s$^{-1}$ lensing a close background source.
An estimate of the redshift of the strongly lensed background source would help to
alleviate the doubt.
We also note that the redshift of this structure has been estimated photometrically.
If this estimation is biased such that this structure is located at lower redshift, then 
we are currently overestimating its velocity dispersion.

\section{Discussion}

In this Section, we discuss the results obtained on the group sample.
Our statements are clearly limited by the large error bars obtained on the
quantities of interest, in particular on the weak-lensing mass estimates.
Therefore, part of the following discussion is qualitative only.

\subsection{Mass is Traced by Light}
The strong lensing system is always found to coincide with the peak of the light
distribution, as can be seen in Fig.~\ref{all} (see, however, the case of
SL2S\,J08544-0121 discussed in the Appendix).
Because a strong lensing system is always associated with a significant mass 
concentration (occurrence of strong lensing implies $\Sigma\,>\,\Sigma_{\rm 
critic}$), this suggests that the light is a good tracer of the mass and
this justifies the choice of the strong lensing system as the group centre when we perform a 
weak lensing analysis.
We note that this statement is limited to the fact that the centres of the luminosity
and mass distributions are located approximately at the same position. We cannot
say much at this point because of the limitations of the weak lensing analysis (Section~6.3).

\subsection{A Homogeneous Sample ?}

The sample is homogeneous in term 
of luminosities ranging between 1 and 4 10$^{12}$\,L$_{\sun}$ 
(Fig.~\ref{reslum2}).
However, Fig.~\ref{all} shows that the luminosity contours have very different 
shapes from one group to another and that all but two (SL2S\,J09413-1100 at 
$z=0.38$ and SL2S\,J02254-0737 at $z=0.51$) have complicated shapes, suggesting 
that these intermediate mass structures may be dynamically young.

\subsection{Evolution with Redshift ?}

We looked at possible evolution of the mass-to-light ratio with redshift in 
Fig.~\ref{msurl}. However, given the large error bars on this quantity, we 
cannot reliably infer any trend. 

\subsection{Comparison to Other Studies}
There have been at least two different estimations of the mass-to-light ratio of
galaxy groups from weak lensing.
The first measurement of the average mass-to-light ratios was performed by 
\citet{hoekstra01} in the CNOC2 survey. They stacked the weak lensing signal
obtained for 50 groups that have been selected on the basis of a dynamical 
analysis of group candidates. 
Their weak lensing detection points towards much less massive systems than the 
one presented in this paper: their best-fit singular isothermal sphere yields a 
velocity dispersion of 274$^{+48}_{-59}$ km\,s$^{-1}$. However, they find a 
mass-to-light ratio of 254 $\pm$ 110\,$h$ (solar units, B band rest frame 
corrected for evolution).

Subsequently, another similar study on a sample of 116 CNOC2 galaxy groups was 
performed by \citet{parker05}. Given more statistics, they were able to split 
the group sample into subsets of poor and rich groups. Their poor galaxy groups 
were found to have an average velocity dispersion of 193 $\pm$ 38 km\,s$^{-1}$ 
and a mass-to-light ratio of 134 $\pm$ 26\,$h$ (solar units, B band rest frame), 
while their rich galaxy groups have a velocity dispersion of 270 $\pm$ 
39 km\,s$^{-1}$ and a mass-to-light ratio of 278 $\pm$ 42\,$h$, thus finding 
evidence for a steep increase in the mass-to-light ratio as a function of mass.

The groups presented in the present work are more massive than the rich galaxy 
group sample from \citet{parker05}. Given the error bars obtained on the 
mass-to-light ratios, it is difficult to reliably compare our findings to those 
from \citet{parker05}. However, considering the 6 groups for which we achieved a 
weak lensing detection, we get a mean velocity dispersion of 623 km\,s$^{-1}$ 
and a mean mass-to-light ratio of 242, consistent with the results of 
\citet{parker05}.
Note that these values are close to mass-to-light ratios of galaxy clusters
\citep[\emph{e.g.}][]{carlberg97,bardeau07}.

Our results suggest that after the steep increase in mass-to-light ratio 
as a function of mass pointed out by \citet{parker05} which seems to occur at 
velocity dispersions around 200\,km\,s$^{-1}$ (or mass scale of order 
$\sim$10$^{13}$ M$_{\sun}$), mass-to-light ratios reach a plateau and do not 
evolve much from the mass scale of groups with velocity dispersions around 
300\,km\,s$^{-1}$ up to rich galaxy clusters with velocity dispersion larger 
than 1\,000\,km\,s$^{-1}$. The group sample studied in this article therefore 
helps to bridge the gap between estimates of the mass-to-light ratio of galaxy 
groups with velocity dispersions in the range 200\,-\,300 km\,s$^{-1}$ and for 
galaxy clusters.\\

\subsection{The Naked Cusp Sample}
Eight galaxy groups reported in this article exhibit typical cusp 
configurations, with a clear tangential arc on one side of the deflector.
For 5 of these groups, we have not been able to find the expected counter image
which is usually located on the other side of the deflector, even when HST 
imaging was available. This incomplete cusp configuration is called 'naked cusp' 
\citep[see, \emph{e.g.}][]{bible,oguri04}.
In two cases, SL2S\,J02215-0647 and SL2S\,J08544-0121, the counter image of the 
cusp is clearly identified on the HST images. In SL2S\,J22133+0048, the
detection of the counter image is tentative only. Moreover, note that the case
of SL2S\,J08544-0121 is a bit special since this cusp configuration is not
located at the centre of the total light distribution, whereas the other cusp
groups are. To summarise, we find that most of the cusp groups are naked.
It may be that the current observations did not allow us to find the counter 
images, but this naked state of a cusp can be linked with lens statistic issue.
If CDM haloes have central density slopes $<$~1.5, \citet{oguri04} predict that 
a significant fraction ($>20$\%) of large separation lenses should have naked 
cusp image configurations.
Therefore, our results could suggest that the groups studied in this work may 
have shallow central density slopes.

\subsection{Other Intermediate Separation Lenses}
\label{cassowary}
We claim this work to be the first representative sample of strong lensing 
systems generated by group scale dark matter haloes, with intermediate image 
separation (in the range $\sim 3\arcsec-\,8\arcsec$).
However, at least five intermediate separation lenses have been reported so far,
even though the authors did not always consider the group-scale haloes as being 
responsible for the image separation, and are interpreting the results in terms 
of massive galactic scale haloes. A remarkable example is the so called Cosmic 
Horseshoe \citep{belokurov07,dye08}, an almost complete
Einstein ring of radius 5$\arcsec$ from a luminous red galaxy at $z=0.44$. 
The authors argued that this is the most massive galaxy lens discovered so far, 
with a mass enclosed within its Einstein radius of 5\,10$^{12}$M$_{\sun}$, 
making this galaxy as massive as the entire Local Group. We rather suggest that 
this lensing system is simply produced by a \emph{group-scale} dark-matter halo,
whose centre is populated by the luminous red galaxy. Indeed, this galaxy is the
brightest object in the group of $\sim$\,26 members as revealed by the SDSS 
photometry. As suggested by the nearly perfect circle outlined by the ring,
no external shear was required in the model of the system, which is not
surprising because the deflector traces the centre of the group halo.
This important discovery led to a more systematic data mining looking for large 
separation lenses in the Sloan Digital Sky Survey: the CAmbridge Sloan Survey Of Wide 
ARcs in the skY
(CASSOWARY\footnote{http://www.ast.cam.ac.uk/research/cassowary/}).
Recently, \citet{belokurov08} reported two new large separation lenses, with 
Einstein radius equal to 4$\arcsec$ and 8$\arcsec$.
More recently, \citet{lin}, in a systematic search for bright arcs in the SDSS, 
reported a strongly lensed $z=2$ galaxy by a deflector whose Einstein Radius 
equals 3.82$\arcsec$. We also mention an intermediate separation lens 
(\R $\sim$ 2.5$\arcsec$) found in the CLASS survey, B2108+213 
\citep{mckean,anu08}.

We expect to find new intermediate separation lenses on completion of the 
CFHTLS, which, in combination with group scale lenses from other promising 
dedicated searches in surveys such as the SDSS\footnote{one month after this 
work was submitted, \citet{kubogroups} reported the finding of five new group
scale lenses in the SDSS}, will increase the statistics of intermediate
separation lenses, thus allowing us to get insights on this intermediate mass
range.

\section{Conclusion}

We have presented a sample of 13 strong lensing features with Einstein radii
between 3$\arcsec$ and 8$\arcsec$, i.e. generated by \emph{galaxy group-scale 
dark-matter haloes.}
This is the first representative sample (not yet complete) of intermediate 
image separations lenses, bridging the gap between galaxy and cluster scales.
Our conclusions are the following:
\begin{enumerate}
\item The strong lensing analysis yields Einstein radii between 2.5$\arcsec$ and 
8$\arcsec$.
\item We have studied the luminosity distribution to infer properties on large 
scales: the strong lenses coincide with the peak of the light distribution; 
most of the luminosity maps have complicated shapes, suggesting dynamically young structures.
\item A weak lensing analysis yields $\sigma_{\rm SIS}\sim$ 500\,km\,s$^{-1}$ 
with large error bars and an upper limit of $\sim$\,900\,km\,s$^{-1}$ for the 
whole sample. This confirms that our sample consists of rich groups.
\item $\sigma_{\rm SIS}$ seems to increase with luminosity but given 
the large error bars this is not statistically significant.
\item The mass-to-light ratio for the sample is found M/L$_i$\,$\sim$ 250 
(solar units, corrected for evolution), with an upper limit of 500.
\end{enumerate}
We note again that our conclusions are limited by the large error bars obtained
on the weak lensing inferred mass estimates.

Complementary observations are ongoing: we are currently analysing multi object 
spectroscopy of group members as well as near infrared WIRCAM imaging.
This sample will be complemented in the near future with other observations
in order to help address issues of structure formation and evolution.

Once the Strong Lensing Legacy Survey will be completed and the selection 
function well understood, we will compare the distribution of observed Einstein 
radii to the predictions from \citet{oguri06}, which may provide an 
interesting test of the $\Lambda$CDM scenario.

\input{table}
\section*{Acknowledgement}
ML acknowledges the Agence Nationale de la Recherche (Project number 06-BLAN-0067)
and the Centre National d' Etudes Spatiales (CNES) for their support.
The Dark Cosmology Centre is funded by the Danish National Research Foundation.
JR is grateful to Caltech for its support.
JPK acknowledges the Centre National de la Recherche Scientifique for its support.
Part of this project is done under the support of National Natural Science
Foundation of China No. 10878003,  10778752, 973Program No. 2007CB815402,
Shanghai Foundation No. 08240514100, 07dz22020, and the Leading Academic
Discipline Project of Shanghai Normal University（DZL805）.
ML acknowledges S\'ebastien Bardeau for making its \textsc{sheartool} package available, and
Masamune Oguri for providing the data used to generate Fig.~1.
We acknowledge the referee for a careful reading of the submitted
paper and constructive suggestions.
This research has made use of the NASA/IPAC Extragalactic Database (NED) which is operated by the Jet 
Propulsion Laboratory, California Institute of Technology, under contract with the National Aeronautics 
and Space Administration.

\bibliographystyle{aa} 
\bibliography{11473}

\appendix
\section{Presentation of each Group}
We dedicate below a paragraph to each group to describe its main properties.
The information are summarised in Table~\ref{res} and illustrated in the
Appendix images.
\subsection{Group Description}

\paragraph{SL2S\,J09013-0158 at $z_{\rm phot}=0.30$ (Fig.~\ref{O2}):}
The luminosity distribution is elongated in the north-south direction.
We report a straight arc between the two main bright central galaxies: 
this is a typical beak-to-beak configuration \citep{beak2beak}. 
This arc is found closer to the northern galaxy ($\sim7\arcsec$) than
the southern one. No HST image is available for this group.
A radio source has been reported by \cite{radio} between the two bright central
galaxies. Given the coordinates and associated errors of this radio source, it 
could be associated with one of the two smaller central galaxies.

\paragraph{SL2S\,J22214-0053 at $z_{\rm spec}=0.334$ (Fig.~\ref{G10}):}
The SDSS provides a redshift measurement for the main galaxy populating the 
deflector of 0.334, and a velocity dispersion of 281\,$\pm$\,45\,km\,s$^{-1}$.
The strong lensing deflector is populated by a single bright galaxy.

\paragraph{SL2S\,J08544-0121 at $z_{\rm spec}=0.351$ (Fig.~\ref{G3}):}
The strong lensing deflector is populated by a single bright galaxy
whose ellipticity and position angle equals 0.3 and 25$\pm$5 degrees, respectively.
The luminosity contours are elongated in the east-west direction and define
a position angle consistent with that of the bright galaxy.
We also find the position angle of the SIE halo ($\sim$\,21 degrees, Section 5) 
to be consistent with that of the bright galaxy.

Note (Fig.~\ref{G3}) that the innermost luminosity contour at 
10$^{7}$ L$_{\sun}$\,kpc$^{-2}$ encompass the SL system but also two bright 
galaxies located $\sim$ 54$\arcsec$ east from the SL system, making this light 
distribution bimodal.
This is the only group for which the luminosity distribution is not clearly
dominated by the lens, making this configuration rather exceptional:
the large Einstein radius ($\sim5\arcsec$) points toward a massive structure 
associated with this lens, but the luminosity distribution is found bimodal.
This suggest a dynamically young structure in the process of formation.
This bright galaxy has a spectroscopic redshift measured from Keck of 0.3514.
We detect two multiply imaged systems: the brightest one is perturbed by
a small satellite galaxy whose redshift is equal to 0.3517 (FORS\,2);
and the outer one is seen on the ACS colour image (Fig.~\ref{G3}).
Note how the northern counter image is found much further ($\sim8\arcsec$) than 
the main arc ($\sim5\arcsec$), suggesting a strong contribution from the 
external shear associated with the host galaxy group.
The HST data brings significant additional information on this lensing feature.

\paragraph{SL2S\,J09413-1100 at $z_{\rm spec}=0.385$ (Fig.~\ref{G2}):}
The luminosity contours look circular. The strong lensing deflector is populated 
by a bright galaxy whose stellar halo is extended and presents
a large ellipticity ($b/a$\,=\,0.59) with a position angle of 74 degrees.
Interestingly, this position angle is found compatible with the orientation of 
the luminosity contours. Note that this is the only group for which the central 
galaxy presents an extended stellar halo.
We measured a spectroscopic redshift for this galaxy using FORS\,2 to be 0.385.
We report a blue arc composed by two merging images north of the deflector,
with its counter image. We have not been able to find another counter image
on the other side of the deflector, even after subtraction of the
galaxy on the HST images. 

\paragraph{SL2S\,J14081+5429 at $z_{\rm spec}=0.416$ (Fig.~\ref{O4}):}
The luminosity distribution is elongated in the north-south direction.
The centre of this group is dominated by three bright galaxies aligned in the
north-south direction.
The brightest one (A, mag$_i$=18.03) is the central one, that is also closest to 
the arc feature. North of A is a galaxy B (mag$_i$=18.37) and south of is a 
galaxy C (mag$_i$=18.46). The SDSS provides a redshift measurement for A 
($z=0.415979$) and for B ($z=0.411022$). A straight arc without any detected 
feature is located between bright central galaxies.

\paragraph{SL2S\,J14300+5546 at $z_{\rm spec}=0.435$ (Fig.~\ref{G6}):}
The luminosity distribution is elongated in the east-west direction.
The strong lensing deflector is populated by a single galaxy whose redshift 
equals 0.435 (Gemini). A tangential arc is found south-east of the deflector.
The position angle of the galaxy populating the deflector ($\sim$\,45 degrees)
is consistent with the position angle defined by the luminosity distribution.

\paragraph{SL2S\,J02140-0532 at $z_{\rm spec}=0.444$ (Fig.~\ref{G1}):}
The luminosity distribution is elongated in the north-south direction.
The strong lensing deflector is populated by three galaxies. The two brightest 
ones have redshift measured spectroscopically from Keck (0.4422 and 0.4439) and 
may be bound gravitationally. We report an arc north of the deflector
composed by two merging images showing substructure. The counter image is easily 
identified east of the deflector. Note that on the ground based image, we see a 
possible counter image south of the deflector which seems to have colour 
compatible with the other images. However, on the ACS data, it is clear that 
this feature cannot be associated with the proposed multiply imaged system. This 
is also confirmed by the modelling: no acceptable fit was able to reproduce the 
multiple images as one could have inferred from the ground based data.
We note, however, that a recent independent study by \citet{alardalone} 
considers the southern image as part of the multiply imaged system. Spectroscopy 
of each feature is needed in order to remove the uncertainty.
The HST data brings significant additional information on this lensing feature:
not only does it help to identify the different images belonging to the same 
system, but it also resolves substructures within each lensed image, increasing 
the number of constraints for the analysis.
The halo of the deflector is oriented with a position angle well constrained at
111$\pm$3 degrees. This is the same direction as the one defined by the luminosity contours.

\paragraph{SL2S\,J02254-0737 at $z_{\rm spec}=0.511$ (Fig.~\ref{G8}):}
The luminosity contours look circular.
The strong lensing deflector is populated by a single galaxy whose redshift 
equals 0.511 (Gemini).
We observe a tangential arc north of the lens galaxy, likely to
be composed by two merging images, with an additional counter image a bit 
further east. The location of the deflector coincide with a radio emission 
reported by \citet{radio}.

\paragraph{SL2S\,J22130-0030 at $z_{\rm phot}=0.61$ (Fig.~\ref{O5}):}
The luminosity distribution is elongated in the north-south direction.
The HST image reveals that the strong lensing deflector is populated by a very 
compact group of at least 6 galaxies.
We report a blue arc east of the deflector, and a likely counter image 
presenting the same colour as the arc on the other side of the deflector.

\paragraph{SL2S\,J02215-0647 at $z_{\rm spec}=0.618$ (Fig.~\ref{G4}):}
We find a gap of 1.3 magnitude in the R band between the brightest and the
second brightest galaxy, not enough to be considered as a fossil group.
The strong lensing deflector is populated by a single galaxy whose redshift
equals 0.618 (FORS\,2).
We report an arc south of the deflector, composed by two merging images.
There is a counter image south-east of the deflector, as well as an additional
counter image located on the other side of the deflector, resolved by HST data.
We report a possible second multiply imaged system constituted by two images
with same CFHTLS colours.

\paragraph{SL2S\,J14314+5533 at $z_{\rm phot}=0.64$ (Fig.~\ref{O3}):}
The luminosity distribution is elongated in the south-east north-west direction.
A small tangential arc is found around bright galaxies.
We cannot conjugate any images that may merge to form the arc.
It is possible that this blue lensed feature in fact is singly imaged.

\paragraph{SL2S\,J08591-0345 at $z_{\rm spec}=0.647$ (Fig.~\ref{O1}):}
The lens is located close (75$\arcsec$\,=\,508\,kpc) to the edge of the field
of view. It is very likely that we are missing a significant part of this group.
Therefore, we have not been reported any values of this group luminosity, 
neither did we pursue any weak lensing analysis for this group.
We observe a rather exotic strong lensing configuration:
the deflector is populated by three bright galaxies and two smaller ones.
One of the bright one has a redshift of 0.647 (FORS\,2).
The multiple images draw an oval contour around the deflector.
It is difficult at this point to know how many multiple images we observe and if 
they are coming from a single or multiple background sources.
An advanced modelling of this exotic lens would be very interesting and will
require deep multi colour space based data with dedicated ground based 
spectroscopy.

\paragraph{SL2S\,J22133+0048 at $z_{\rm phot}=0.83$ (Fig.~\ref{G9}):}
The luminosity distribution is elongated in the south-east north-west direction.
The strong lensing deflector is populated by a single galaxy.
A tangential arc composed by two merging images is found east of the deflector.
A possible counter image on the other side of the deflector is detected on the
space based images.
The location of the deflector coincides with a radio emission reported by 
\citet{radio}.
Our weak lensing analysis yields a velocity dispersion between 432 and 1218 km\,s$^{-1}$.
The upper limit is therefore consistent with a galaxy cluster.
Since its Einstein radius is estimated to be about 3.4$\arcsec$, it is possible that SL2S\,J22133+0048 
is indeed a galaxy cluster with a velocity dispersion
larger than 1000 km\,s$^{-1}$ lensing a close background source.
An estimate of the redshift of the strongly lensed background source would help to
alleviate the doubt.
We note also that the redshift of this structure has been estimated photometrically.
If this estimation is biased and if this structure is located at lower redshift, then 
we are currently overestimating its velocity dispersion.

\end{document}

%% file: table.tex
\hspace{-2cm}
\begin{sidewaystable*}
\begin{center}
\begin{tabular}{lcccccccccccccc}
\hline
SL2S Group & R.A. & Decl. & $z_{\rm spec}$ & $z_{\rm phot}$ & R$_{\textsc{e}}$ (\textsc{sl}) & R$_{\textsc{e}}$ (\textsc{wl}) & D$_{\rm ls}$/D$_{\rm s}$ & $z_{\rm eff}$ & $\sigma_{\rm SIS}$ (\textsc{wl}) & M$_{\textsc{wl}}$\,(2\,Mpc) & N$_{\rm bckg}$ & L & Seeing & M$_{\rm c}$ \\
\hline
J09013-0158 &  135.41443 &  -1.9811666 & - & 0.296$^{+0.052}_{-0.017}$ &$\sim$ 6.8 & 3.83$\pm$2.97 & 0.487 & 0.641 & 521$^{+174}_{-274}$ & 4.0$\pm$3.1 (3.9$\pm$3.0) & 11.8 & 2.78$\pm$0.38 & 0.80 &24.13 \\
\hline
J22214-0053 & 335.43226 & -0.88404303 &  0.334  & 0.314$^{+0.026}_{-0.057}$ & 6.83$\pm$0.15 & 3.15$\pm$3.10 & 0.448  & 0.660 & 494$^{+201}_{-432}$ & 3.6$\pm$3.5 (3.5$\pm$3.4) & 10.3 & 1.36$\pm$0.27 & 0.72  &  24.60 \\
\hline
J08544-0121 & 133.69395 &  -1.3603506 & 0.351 & 0.324$^{+0.067}_{-0.013}$ & 5.48$\pm$0.12 &  5.37$\pm$2.12 & 0.430 & 0.671 & 658$^{+119}_{-146}$ & 6.3$\pm$2.5 (6.2$\pm$2.4) & 13.5 & 3.29$\pm$0.40 & 0.51 & 23.91\\
\hline
J09413-1100 & 145.39478 & -11.015091 & 0.385 &  0.418$^{+0.007}_{-0.005}$ &7.50$\pm$0.81 &  2.98$\pm$2.64 & 0.393 & 0.688 & 508$^{+189}_{-337}$  & 3.7$\pm$3.4 (3.7$\pm$3.3) & 10.4 & 1.27$\pm$0.24 & 0.69 & 24.25 \\
\hline
J14081+5429 & 212.05808 & 54.484634 & 0.416 &  0.436$^{+0.016}_{-0.039}$ & $\sim$ 4.5 & $<$ 4.83 & 0.362 & 0.705 & $<$ 680 & $<$ 6.8 (6.6) & 10.8  & 2.46$\pm$0.39 & 0.72 & 23.61 \\
\hline
J14300+5546 & 217.50275 & 55.779964  & 0.435 &  0.486$^{+0.028}_{-0.024}$ &4.69$\pm$1.43 & $<$ 4.64 & 0.344 & 0.715 & $<$ 684 & $<$ 6.8 (6.6) & 14.7 & 2.15$\pm$0.27 & 0.72 & 24.39 \\
\hline
J02140-0532& 33.533465 & -5.5923947  & 0.444 & 0.440$^{+0.026}_{-0.016}$ & 7.31$\pm$0.49 & 3.62$\pm$2.45& 0.335 & 0.719 & 612$^{+180}_{-264}$ & 5.5$\pm$3.7 (5.3$\pm$3.6) & 13.0 &  2.90$\pm$0.39 & 0.61 & 23.96 \\
\hline
J02254-0737 & 36.44216 & -7.6273523 & 0.511 & 0.480$^{+0.016}_{-0.016}$ & 6.00$\pm$0.31 & $<$ 2.56 & 0.274 & 0.751 & $<$ 569 & $<$ 4.7 (4.4) & 11.9  & 1.93$\pm$0.30 & 0.67 & 23.76  \\
\hline
J22130-0030 & 333.27883 & -0.51030426 & - & 0.606$^{+0.012}_{-0.029}$ & $\sim$ 3.6 & $<$ 3.06 & 0.201 & 0.805 & $<$ 726 & $<$ 6.9 (6.2) & 11.8 & 3.19$\pm$0.50 & 0.61 & 24.41 \\
\hline
J02215-0647 & 35.463263 & -6.792443 &  0.618 & 0.636$^{+0.028}_{-0.044}$ & 2.56$\pm$0.14  & $<$ 1.22 & 0.196 & 0.810 & $<$ 463 & $<$ 3.1 (2.8) & 12.0 & 1.50$\pm$0.27 & 0.51 & 24.24 \\
\hline
J14314+5533 & 217.9156 & 55.556308 & - & 0.640$^{+0.143}_{-0.069}$ & $\sim$ 2.5 & $<$ 3.98 & 0.182 &0.822 & $<$ 870 & $<$ 11.1 (9.9) & 13.4 & 3.55$\pm$0.48 & 0.72 & 24.14 \\
\hline
J08591-0345 & 134.81024 &  -3.7540217 &  0.647 & 0.586$^{+0.040}_{-0.016}$ & $\sim$ 8 & - & - & - & -  & - & - &-& 0.82 & 23.79 \\
\hline
J22133+0048 & 333.38274 &  0.81002713 & - & 0.826$^{+0.022}_{-0.032}$ & 3.40$\pm$0.2  & 2.46$\pm$1.91 & 0.095 & 0.946 & 947$^{+315}_{-499}$ & 13.1$\pm$10.2 (10.8$\pm$8.5) & 11.1 &  3.34$\pm$0.54 & 0.54 & 24.08 \\
\hline
\end{tabular}
\end{center}
\caption{Summary of the main properties of SL2S groups
{\bf{J\,2000.0 coordinates of the main galaxy populating the deflector;
spectroscopic redshift $z_{\rm spec}$; photometric redshift $z_{\rm phot}$; Einstein radius derived from the strong lensing modelling, when possible.
For the non cusp group, we report with a \~\ the distance between the lensed feature and the brighest
group member;
Einstein radius derived from fitting a Singular Isothermal Sphere (SIS) to the weak lensing signal;
mean geometrical factor D$_{\rm ls}$/D$_{\rm s}$ and corresponding effective redshift $z_{\rm eff}$; velocity dispersion;
projected mass derived from weak lensing computed within a circular aperture of radius equals to 2\,Mpc, in units 10$^{14}$ M$_{\sun}$
(masses derived using the \citet{coupon} photometric redshift catalogue are given within brackets);
density of background galaxies, expressed in objects per square arcminute;
total $i$ band rest frame luminosity corrected for evolution, in units 10$^{12}$ L$_{\sun}$;
seeing of the $i$ band observations; $i$ band completeness magnitude.
Einstein radii and seeings are expressed in arcseconds.
}}}
\label{res}
\end{sidewaystable*}

%% file: 11473.bbl
\begin{thebibliography}{74}
\expandafter\ifx\csname natexlab\endcsname\relax\def\natexlab#1{#1}\fi

\bibitem[{{Alard}(2009)}]{alardalone}
{Alard}, C. 2009, ArXiv e-prints 0901.0344

\bibitem[{{Auger} {et~al.}(2007){Auger}, {Fassnacht}, {Abrahamse}, {Lubin}, \&
  {Squires}}]{auger07}
{Auger}, M.~W., {Fassnacht}, C.~D., {Abrahamse}, A.~L., {Lubin}, L.~M., \&
  {Squires}, G.~K. 2007, \aj, 134, 668

\bibitem[{{Auger} {et~al.}(2008){Auger}, {Fassnacht}, {Wong}, {Thompson},
  {Matthews}, \& {Soifer}}]{auger08}
{Auger}, M.~W., {Fassnacht}, C.~D., {Wong}, K.~C., {et~al.} 2008, \apj, 673,
  778

\bibitem[{{Bardeau}(2004)}]{sebphd}
{Bardeau}, S. 2004, PhD thesis, Universit\'e Paul Sabatier, Toulouse III,
  France

\bibitem[{{Bardeau} {et~al.}(2005){Bardeau}, {Kneib}, {Czoske}, {Soucail},
  {Smail}, {Ebeling}, \& {Smith}}]{bardeau05}
{Bardeau}, S., {Kneib}, J.-P., {Czoske}, O., {et~al.} 2005, \aap, 434, 433

\bibitem[{{Bardeau} {et~al.}(2007){Bardeau}, {Soucail}, {Kneib}, {Czoske},
  {Ebeling}, {Hudelot}, {Smail}, \& {Smith}}]{bardeau07}
{Bardeau}, S., {Soucail}, G., {Kneib}, J.-P., {et~al.} 2007, \aap, 470, 449

\bibitem[{{Belokurov} {et~al.}(2009){Belokurov}, {Evans}, {Hewett}, {Moiseev},
  {McMahon}, {Sanchez}, \& {King}}]{belokurov08}
{Belokurov}, V., {Evans}, N.~W., {Hewett}, P.~C., {et~al.} 2009, \mnras, 392,
  104

\bibitem[{{Belokurov} {et~al.}(2007){Belokurov}, {Evans}, {Moiseev}, {King},
  {Hewett}, {Pettini}, {Wyrzykowski}, {McMahon}, {Smith}, {Gilmore}, {Sanchez},
  {Udalski}, {Koposov}, {Zucker}, \& {Walcher}}]{belokurov07}
{Belokurov}, V., {Evans}, N.~W., {Moiseev}, A., {et~al.} 2007, \apjl, 671, L9

\bibitem[{{Bertin} \& {Arnouts}(1996)}]{sextractor}
{Bertin}, E. \& {Arnouts}, S. 1996, \aap, 117, 393

\bibitem[{{Bolzonella} {et~al.}(2000){Bolzonella}, {Miralles}, \& {Pell{\'
  o}}}]{hyperz}
{Bolzonella}, M., {Miralles}, J.-M., \& {Pell{\' o}}, R. 2000, \aap, 363, 476

\bibitem[{{Bridle} {et~al.}(2002){Bridle}, {Kneib}, {Bardeau}, \&
  {Gull}}]{im2shape}
{Bridle}, S., {Kneib}, J.-P., {Bardeau}, S., \& {Gull}, S. 2002, in The shapes
  of galaxies and their dark halos, Proceedings of the Yale Cosmology Workshop
  , New Haven, Connecticut, USA, 28-30 May 2001. Edited by Priyamvada
  Natarajan., ed. P.~{Natarajan}, 38--+

\bibitem[{{Bruzual} \& {Charlot}(2003)}]{BC03}
{Bruzual}, G. \& {Charlot}, S. 2003, \mnras, 344, 1000

\bibitem[{{Cabanac} {et~al.}(2007){Cabanac}, {Alard}, {Dantel-Fort}, {Fort},
  {Gavazzi}, {Gomez}, {Kneib}, {Le F{\`e}vre}, {Mellier}, {Pello}, {Soucail},
  {Sygnet}, \& {Valls-Gabaud}}]{sl2s}
{Cabanac}, R.~A., {Alard}, C., {Dantel-Fort}, M., {et~al.} 2007, \aap, 461, 813

\bibitem[{{Carlberg} {et~al.}(1997){Carlberg}, {Yee}, \&
  {Ellingson}}]{carlberg97}
{Carlberg}, R.~G., {Yee}, H.~K.~C., \& {Ellingson}, E. 1997, \apj, 478, 462

\bibitem[{{Carrasco} {et~al.}(2007){Carrasco}, {Cypriano}, {Neto}, {Cuevas},
  {Sodr{\'e}}, {de Oliveira}, \& {Ramirez}}]{edu07}
{Carrasco}, E.~R., {Cypriano}, E.~S., {Neto}, G.~B.~L., {et~al.} 2007, \apj,
  664, 777

\bibitem[{{Condon} {et~al.}(1998){Condon}, {Cotton}, {Greisen}, {Yin},
  {Perley}, {Taylor}, \& {Broderick}}]{radio}
{Condon}, J.~J., {Cotton}, W.~D., {Greisen}, E.~W., {et~al.} 1998, \aj, 115,
  1693

\bibitem[{{Coupon} {et~al.}(2008){Coupon}, {Ilbert}, {Kilbinger}, {McCracken},
  {Mellier}, {Arnouts}, {Bertin}, {Hudelot}, {Schultheis}, {Le F{\`e}vre}, {Le
  Brun}, {Guzzo}, {Bardelli}, {Zucca}, {Bolzonella}, {Garilli}, {Zamorani}, \&
  {Zanichelli}}]{coupon}
{Coupon}, J., {Ilbert}, O., {Kilbinger}, M., {et~al.} 2008, ArXiv e-prints
  0811.3326

\bibitem[{{Cypriano} {et~al.}(2005){Cypriano}, {Lima Neto}, {Sodr{\'e}},
  {Kneib}, \& {Campusano}}]{edu05}
{Cypriano}, E.~S., {Lima Neto}, G.~B., {Sodr{\'e}}, Jr., L., {Kneib}, J.-P., \&
  {Campusano}, L.~E. 2005, \apj, 630, 38

\bibitem[{{Cypriano} {et~al.}(2004){Cypriano}, {Sodr{\'e}}, {Kneib}, \&
  {Campusano}}]{edu}
{Cypriano}, E.~S., {Sodr{\'e}}, L.~J., {Kneib}, J.-P., \& {Campusano}, L.~E.
  2004, \apj, 613, 95

\bibitem[{{D'Onghia} {et~al.}(2005){D'Onghia}, {Sommer-Larsen}, {Romeo},
  {Burkert}, {Pedersen}, {Portinari}, \& {Rasmussen}}]{elena05}
{D'Onghia}, E., {Sommer-Larsen}, J., {Romeo}, A.~D., {et~al.} 2005, \apjl, 630,
  L109

\bibitem[{{Dye} {et~al.}(2008){Dye}, {Evans}, {Belokurov}, {Warren}, \&
  {Hewett}}]{dye08}
{Dye}, S., {Evans}, N.~W., {Belokurov}, V., {Warren}, S.~J., \& {Hewett}, P.
  2008, \mnras, 388, 384

\bibitem[{{Faltenbacher} \& {Mathews}(2007)}]{group7}
{Faltenbacher}, A. \& {Mathews}, W.~G. 2007, \mnras, 375, 313

\bibitem[{{Fassnacht} \& {Lubin}(2002)}]{fassnacht02}
{Fassnacht}, C.~D. \& {Lubin}, L.~M. 2002, \aj, 123, 627

\bibitem[{{Fassnacht} {et~al.}(2006){Fassnacht}, {McKean}, {Koopmans}, {Treu},
  {Blandford}, {Auger}, {Jeltema}, {Lubin}, {Margoniner}, \&
  {Wittman}}]{fassnacht06}
{Fassnacht}, C.~D., {McKean}, J.~P., {Koopmans}, L.~V.~E., {et~al.} 2006, \apj,
  651, 667

\bibitem[{{Faure} {et~al.}(2008){Faure}, {Kneib}, {Hilbert}, {Massey},
  {Covone}, {Finoguenov}, {Leauthaud}, {Taylor}, {Pires}, \&
  {Scoville}}]{cecile}
{Faure}, C., {Kneib}, J.~., {Hilbert}, S., {et~al.} 2008, ArXiv e-prints
  0810.4838

\bibitem[{{Feron} {et~al.}(2008){Feron}, {Hjorth}, {McKean}, \&
  {Samsing}}]{chloe}
{Feron}, C., {Hjorth}, J., {McKean}, J.~P., \& {Samsing}, J. 2008, ArXiv
  e-prints 0810.0780

\bibitem[{{Finoguenov} {et~al.}(2007){Finoguenov}, {Ponman}, {Osmond}, \&
  {Zimer}}]{group5}
{Finoguenov}, A., {Ponman}, T.~J., {Osmond}, J.~P.~F., \& {Zimer}, M. 2007,
  \mnras, 374, 737

\bibitem[{{Gastaldello} {et~al.}(2007){Gastaldello}, {Buote}, {Humphrey},
  {Zappacosta}, {Bullock}, {Brighenti}, \& {Mathews}}]{fabio}
{Gastaldello}, F., {Buote}, D.~A., {Humphrey}, P.~J., {et~al.} 2007, \apj, 669,
  158

\bibitem[{{Gladders} {et~al.}(1998){Gladders}, {Lopez-Cruz}, {Yee}, \&
  {Kodama}}]{gladders}
{Gladders}, M.~D., {Lopez-Cruz}, O., {Yee}, H.~K.~C., \& {Kodama}, T. 1998,
  \apj, 501, 571

\bibitem[{{Grillo} {et~al.}(2008){Grillo}, {Lombardi}, {Rosati}, {Bertin},
  {Gobat}, {Demarco}, {Lidman}, {Motta}, \& {Nonino}}]{grillo12}
{Grillo}, C., {Lombardi}, M., {Rosati}, P., {et~al.} 2008, \aap, 486, 45

\bibitem[{{Helsdon} \& {Ponman}(2000)}]{group1}
{Helsdon}, S.~F. \& {Ponman}, T.~J. 2000, \mnras, 315, 356

\bibitem[{{Helsdon} \& {Ponman}(2003)}]{group2}
{Helsdon}, S.~F. \& {Ponman}, T.~J. 2003, \mnras, 339, L29

\bibitem[{{Heymans} {et~al.}(2006){Heymans}, {Van Waerbeke}, {Bacon}, {Berge},
  {Bernstein}, {Bertin}, {Bridle}, {Brown}, {Clowe}, {Dahle}, {Erben}, {Gray},
  {Hetterscheidt}, {Hoekstra}, {Hudelot}, {Jarvis}, {Kuijken}, {Margoniner},
  {Massey}, {Mellier}, {Nakajima}, {Refregier}, {Rhodes}, {Schrabback}, \&
  {Wittman}}]{step1}
{Heymans}, C., {Van Waerbeke}, L., {Bacon}, D., {et~al.} 2006, \mnras, 368,
  1323

\bibitem[{{Hoekstra}(2007)}]{hoekstrasample}
{Hoekstra}, H. 2007, \mnras, 379, 317

\bibitem[{{Hoekstra} {et~al.}(2001){Hoekstra}, {Franx}, {Kuijken}, {Carlberg},
  {Yee}, {Lin}, {Morris}, {Hall}, {Patton}, {Sawicki}, \& {Wirth}}]{hoekstra01}
{Hoekstra}, H., {Franx}, M., {Kuijken}, K., {et~al.} 2001, \apjl, 548, L5

\bibitem[{{Ienna} \& {Pell{\'o}}(2006)}]{iena}
{Ienna}, F. \& {Pell{\'o}}, R. 2006, in SF2A-2006: Semaine de l'Astrophysique
  Francaise, ed. D.~{Barret}, F.~{Casoli}, G.~{Lagache}, A.~{Lecavelier}, \&
  L.~{Pagani}, 347--+

\bibitem[{{Inada} {et~al.}(2008){Inada}, {Oguri}, {Shin}, {Kayo}, {Strauss},
  {Morokuma}, {Schneider}, {Becker}, {Bahcall}, \& {York}}]{inada}
{Inada}, N., {Oguri}, M., {Shin}, M.-S., {et~al.} 2008, ArXiv e-prints
  0809.0912

\bibitem[{{Jing} \& {Suto}(2002)}]{jingsuto}
{Jing}, Y.~P. \& {Suto}, Y. 2002, \apj, 574, 538

\bibitem[{{Jullo} {et~al.}(2007){Jullo}, {Kneib}, {Limousin},
  {El{\'{\i}}asd{\'o}ttir}, {Marshall}, \& {Verdugo}}]{jullo07}
{Jullo}, E., {Kneib}, J.-P., {Limousin}, M., {et~al.} 2007, New Journal of
  Physics, 9, 447

\bibitem[{{Kassiola} {et~al.}(1992){Kassiola}, {Kovner}, \&
  {Blandford}}]{beak2beak}
{Kassiola}, A., {Kovner}, I., \& {Blandford}, R.~D. 1992, \apj, 396, 10

\bibitem[{{King}(2007)}]{king1689}
{King}, L.~J. 2007, \mnras, 956

\bibitem[{{Kneib} {et~al.}(2003){Kneib}, {Hudelot}, {Ellis}, {Treu}, {Smith},
  {Marshall}, {Czoske}, {Smail}, \& {Natarajan}}]{kneib03}
{Kneib}, J., {Hudelot}, P., {Ellis}, R.~S., {et~al.} 2003, \apj, 598, 804

\bibitem[{{Kovner}(1987)}]{kovner}
{Kovner}, I. 1987, \apj, 321, 686

\bibitem[{{Kubo} {et~al.}(2008){Kubo}, {Allam}, {Annis}, {Buckley-Geer},
  {Diehl}, {Kubik}, {Lin}, \& {Tucker}}]{kubogroups}
{Kubo}, J.~M., {Allam}, S.~S., {Annis}, J., {et~al.} 2008, ArXiv e-prints
  0812.3934

\bibitem[{{Kundic} {et~al.}(1997){Kundic}, {Hogg}, {Blandford}, {Cohen},
  {Lubin}, \& {Larkin}}]{kundic97a}
{Kundic}, T., {Hogg}, D.~W., {Blandford}, R.~D., {et~al.} 1997, \aj, 114, 2276

\bibitem[{{Limousin} {et~al.}(2007{\natexlab{a}}){Limousin}, {Kneib},
  {Bardeau}, {Natarajan}, {Czoske}, {Smail}, {Ebeling}, \& {Smith}}]{mypaperII}
{Limousin}, M., {Kneib}, J.~P., {Bardeau}, S., {et~al.} 2007{\natexlab{a}},
  \aap, 461, 881

\bibitem[{{Limousin} {et~al.}(2007{\natexlab{b}}){Limousin}, {Richard},
  {Jullo}, {Kneib}, {Fort}, {Soucail}, {El{\'{\i}}asd{\'o}ttir}, {Natarajan},
  {Ellis}, {Smail}, {Czoske}, {Smith}, {Hudelot}, {Bardeau}, {Ebeling},
  {Egami}, \& {Knudsen}}]{mypaperIII}
{Limousin}, M., {Richard}, J., {Jullo}, E., {et~al.} 2007{\natexlab{b}}, \apj,
  668, 643

\bibitem[{{Lin} {et~al.}(2008){Lin}, {Buckley-Geer}, {Allam}, {Tucker},
  {Diehl}, {Kubik}, {Kubo}, {Annis}, {Frieman}, {Oguri}, \& {Inada}}]{lin}
{Lin}, H., {Buckley-Geer}, E., {Allam}, S.~S., {et~al.} 2008, ArXiv e-prints
  0809.4475

\bibitem[{{Mamon}(2007)}]{mamongroup}
{Mamon}, G.~A. 2007, in Groups of Galaxies in the Nearby Universe, ed.
  I.~{Saviane}, V.~D. {Ivanov}, \& J.~{Borissova}, 203--+

\bibitem[{{Mandelbaum} {et~al.}(2006){Mandelbaum}, {Seljak}, {Cool}, {Blanton},
  {Hirata}, \& {Brinkmann}}]{rachelgroup}
{Mandelbaum}, R., {Seljak}, U., {Cool}, R.~J., {et~al.} 2006, \mnras, 372, 758

\bibitem[{{McKean} {et~al.}(2005){McKean}, {Browne}, {Jackson}, {Koopmans},
  {Norbury}, {Treu}, {York}, {Biggs}, {Blandford}, {de Bruyn}, {Fassnacht},
  {Mao}, {Myers}, {Pearson}, {Phillips}, {Readhead}, {Rusin}, \&
  {Wilkinson}}]{mckean}
{McKean}, J.~P., {Browne}, I.~W.~A., {Jackson}, N.~J., {et~al.} 2005, \mnras,
  356, 1009

\bibitem[{{Medezinski} {et~al.}(2007){Medezinski}, {Broadhurst}, {Umetsu},
  {Coe}, {Ben{\'{\i}}tez}, {Ford}, {Rephaeli}, {Arimoto}, \& {Kong}}]{elinor}
{Medezinski}, E., {Broadhurst}, T., {Umetsu}, K., {et~al.} 2007, \apj, 663, 717

\bibitem[{{Mellier}(1999)}]{mellier99}
{Mellier}, Y. 1999, \araa, 37, 127

\bibitem[{{M{\"o}ller} {et~al.}(2002){M{\"o}ller}, {Natarajan}, {Kneib}, \&
  {Blain}}]{ole}
{M{\"o}ller}, O., {Natarajan}, P., {Kneib}, J.-P., \& {Blain}, A.~W. 2002,
  \apj, 573, 562

\bibitem[{{Momcheva} {et~al.}(2006){Momcheva}, {Williams}, {Keeton}, \&
  {Zabludoff}}]{momcheva}
{Momcheva}, I., {Williams}, K., {Keeton}, C., \& {Zabludoff}, A. 2006, \apj,
  641, 169

\bibitem[{{More} {et~al.}(2008){More}, {McKean}, {Muxlow}, {Porcas},
  {Fassnacht}, \& {Koopmans}}]{anu08}
{More}, A., {McKean}, J.~P., {Muxlow}, T.~W.~B., {et~al.} 2008, \mnras, 384,
  1701

\bibitem[{{Morgan} {et~al.}(2005){Morgan}, {Kochanek}, {Pevunova}, \&
  {Schechter}}]{morgan05}
{Morgan}, N.~D., {Kochanek}, C.~S., {Pevunova}, O., \& {Schechter}, P.~L. 2005,
  \aj, 129, 2531

\bibitem[{{Newton} {et~al.}(2008){Newton}, {Marshall}, \&
  {Treu}}]{slacsclustering}
{Newton}, E.~R., {Marshall}, P.~J., \& {Treu}, T. 2008, ArXiv e-prints
  0810.3934

\bibitem[{{Oguri}(2006)}]{oguri06}
{Oguri}, M. 2006, \mnras, 367, 1241

\bibitem[{{Oguri} \& {Keeton}(2004)}]{oguri04}
{Oguri}, M. \& {Keeton}, C.~R. 2004, \apj, 610, 663

\bibitem[{{Oguri} {et~al.}(2005){Oguri}, {Keeton}, \& {Dalal}}]{oguri05}
{Oguri}, M., {Keeton}, C.~R., \& {Dalal}, N. 2005, \mnras, 364, 1451

\bibitem[{{Osmond} \& {Ponman}(2004)}]{group3}
{Osmond}, J.~P.~F. \& {Ponman}, T.~J. 2004, \mnras, 350, 1511

\bibitem[{{Parker} {et~al.}(2005){Parker}, {Hudson}, {Carlberg}, \&
  {Hoekstra}}]{parker05}
{Parker}, L.~C., {Hudson}, M.~J., {Carlberg}, R.~G., \& {Hoekstra}, H. 2005,
  \apj, 634, 806

\bibitem[{{Rasmussen} \& {Ponman}(2007)}]{group6}
{Rasmussen}, J. \& {Ponman}, T.~J. 2007, \mnras, 380, 1554

\bibitem[{{Schneider} {et~al.}(1992){Schneider}, {Ehlers}, \& {Falco}}]{bible}
{Schneider}, P., {Ehlers}, J., \& {Falco}, E.~E. 1992, {Gravitational Lenses}
  (Berlin: Springer-Verlag)

\bibitem[{{Sommer-Larsen}(2006)}]{jespergroup}
{Sommer-Larsen}, J. 2006, \mnras, 369, 958

\bibitem[{{Sun} {et~al.}(2009){Sun}, {Voit}, {Donahue}, {Jones}, {Forman}, \&
  {Vikhlinin}}]{sun}
{Sun}, M., {Voit}, G.~M., {Donahue}, M., {et~al.} 2009, \apj, 693, 1142

\bibitem[{{Treu} {et~al.}(2009){Treu}, {Gavazzi}, {Gorecki}, {Marshall},
  {Koopmans}, {Bolton}, {Moustakas}, \& {Burles}}]{treuaroundslacs}
{Treu}, T., {Gavazzi}, R., {Gorecki}, A., {et~al.} 2009, \apj, 690, 670

\bibitem[{{Trott} {et~al.}(2008){Trott}, {Treu}, {Koopmans}, \&
  {Webster}}]{trott}
{Trott}, C.~M., {Treu}, T., {Koopmans}, L.~V.~E., \& {Webster}, R.~L. 2008,
  ArXiv e-prints 0812.0748

\bibitem[{{Tu} {et~al.}(2008){Tu}, {Limousin}, {Fort}, {Shu}, {Sygnet},
  {Jullo}, {Kneib}, \& {Richard}}]{ring1689}
{Tu}, H., {Limousin}, M., {Fort}, B., {et~al.} 2008, \mnras, 1169

\bibitem[{{van den Bosch} {et~al.}(2008){van den Bosch}, {Pasquali}, {Yang},
  {Mo}, {Weinmann}, {McIntosh}, \& {Aquino}}]{group9}
{van den Bosch}, F.~C., {Pasquali}, A., {Yang}, X., {et~al.} 2008, ArXiv
  e-prints 0805.002

\bibitem[{{Williams} {et~al.}(2006){Williams}, {Momcheva}, {Keeton},
  {Zabludoff}, \& {Leh{\'a}r}}]{williams06}
{Williams}, K.~A., {Momcheva}, I., {Keeton}, C.~R., {Zabludoff}, A.~I., \&
  {Leh{\'a}r}, J. 2006, \apj, 646, 85

\bibitem[{{Willis} {et~al.}(2005){Willis}, {Pacaud}, {Valtchanov}, {Pierre},
  {Ponman}, {Read}, {Andreon}, {Altieri}, {Quintana}, {Dos Santos},
  {Birkinshaw}, {Bremer}, {Duc}, {Galaz}, {Gosset}, {Jones}, \&
  {Surdej}}]{group4}
{Willis}, J.~P., {Pacaud}, F., {Valtchanov}, I., {et~al.} 2005, \mnras, 363,
  675

\bibitem[{{Yang} {et~al.}(2008){Yang}, {Mo}, \& {van den Bosch}}]{group8}
{Yang}, X., {Mo}, H.~J., \& {van den Bosch}, F.~C. 2008, \apj, 676, 248

\end{thebibliography}
